\documentclass[aps,prl,reprint,superscriptaddress]{revtex4-1}

\usepackage{graphicx}
\usepackage{braket}
\usepackage{amsmath,amssymb,amsfonts}
\usepackage{hyperref}

\usepackage{enumitem}

\usepackage{xcolor}
\usepackage{soul,xcolor}
\setstcolor{red}
\definecolor{OliveGreen}{rgb}{0,0.5,0}

\newcommand{\fref}[1]{Fig.~\ref{#1}}

\begin{document}

\title{Oscillations and decay of superfluid currents in a one-dimensional Bose gas on a ring}

\author{Juan Polo}
\affiliation{Univ. Grenoble Alpes, CNRS, LPMMC, 38000 Grenoble, France}
\affiliation{Quantum Systems Unit, Okinawa Institute of Science and Technology Graduate University, Onna, Okinawa 904-0495, Japan}
\author{Romain Dubessy}
\author{Paolo Pedri}
\author{H\'el\`ene Perrin}
\affiliation{Laboratoire de physique des lasers, CNRS, Universit\'e Paris 13,
Sorbonne Paris Cit\'e, 99 avenue J.-B. Cl\'ement, F-93430 Villetaneuse, France}
\author{Anna Minguzzi$^{1}$}

\date{\today}

\begin{abstract}
We study the time evolution of a supercurrent imprinted on a one-dimensional ring of interacting bosons in the presence of a defect created by a localized barrier. Depending on interaction strength and temperature, we identify various dynamical regimes where the current oscillates, is self-trapped or decays with time. We show that the dynamics are captured by a dual Josephson model and  involve phase slips of thermal or quantum nature.  
\end{abstract}


\maketitle
Superfluidity is a fascinating phenomenon emerging in interacting quantum systems and governing their low temperature transport properties. Supercurrents, named in analogy with superconductivity, are characterized, among others, by frictionless flow and quantized vortices, and are most easily evidenced in ring geometries.   
Ultra-cold atoms confined in ring traps have proven to be a great tool to study superfluid transport properties \cite{Ramanathan2011,Moulder2012,Husmann_2015}.
Due to their tunability and their high degree of control, they are an ideal system for studying  the effect of interactions and dimensionality in the superfluid transport dynamics. 
As  superconducting SQUIDs have provided a wealth of applications, the realization of  their atomic analogs -- the AQUID \cite{Mathey_2016} -- is an important step in the field of atomtronics \cite{Seaman_2007,Amico2014,Amico_2017,Gauthier2019}.

From a  fundamental point of view, an open question is the stability of supercurrents. This is related, but complementary to the study of setting the superfluid in rotation, also related to vortex nucleation \cite{Madison2001,Penckwitt2002,Lobo2004}.
For a three-dimensional (3D) ring geometry the stochastic decay of the quantized current has been studied, evidencing the role of the critical velocity \cite{Moulder2012,Dubessy2012}. In the presence of a repulsive barrier crossing the ring, resulting in a weak link, hysteresis in the phase slips dynamics has been investigated \cite{Wright2013a,Wright2013,Eckel2014,Yakimenko_2015,Munoz2015} and the role of thermal activation evidenced \cite{Kumar2017a}.
A scenario for the phase slips dynamics induced by a weak link based on the role of vortices can be used to explain qualitatively the experimental observations \cite{Piazza2009} but fails to account quantitatively for the thermal activation \cite{Mathey2014,Kunimi2017}. Also in a  3D fermionic double-well Josephson junction phase-slips  play a role in the dynamics \cite{Burchianti2018,Xhani2019}.

In this context one question naturally arises: if the phase slips dynamics are driven in 3D by vortices crossing the weak link, what happens in lower dimensions? While in two-dimensional (2D) systems vortices still play a crucial role in the superfluid dynamics \cite{Piazza2009,Mathey_2016}, they cannot exist in one-dimension (1D). Therefore the phase slips phenomenon should be of a different nature in 1D.

Previous works have shown the role of phase-slips~\cite{Danshita2012a,DErrico2017} in the decay of 1D transport in the presence of periodic perturbation~\cite{Tanzi2016}. For a microscopic impurity the decay rate has been estimated by computing the drag force~\cite{Cherny2009}. For sufficiently small obstacles stationary circulating states may exist~\cite{Carr2000a,Cominotti2014,Shamriz_2018}, while a forced flow past a larger obstacle results in soliton emission~\cite{Hakim1997,Freire1997,Katsimiga_2018}. 
Most of the previous studies were performed in a rotating frame, thus \emph{imposing} a flow onto the ring, allowing to estimate the nucleation rate of phase-slips~\cite{Khlebnikov2005a}. For intermediate to strong interactions and small barriers it has been shown that the decay of persistent currents is related to the low-energy  excitations in the ring \cite{Polo_2018}. 

\begin{figure}[t!]
\includegraphics[width=\linewidth]{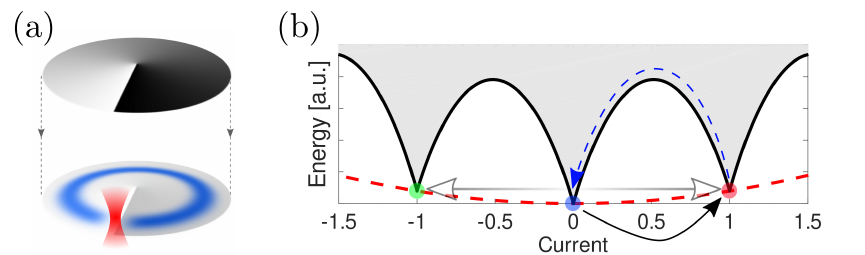}%
\caption{\label{fig:sketch}(a) Sketch of the quench protocol: a 1D Bose gas on a ring in presence of a localized barrier, e.g. a tightly focused repulsive optical potential (red), creating a dip in the density (blue) is quenched out of equilibrium by phase imprinting.  (b) Energy landscape of the homogeneous 1D Bose gas on a ring: the states with integer values of the current per particle correspond to local minima of the energy. The quench (black arrow) transfers the system from the initial zero-current state (light blue circle) to the state with one unit of current (light red circle). Depending on the parameters, the barrier can resonantly couple the +1 and -1 states (light gray arrow) or induce an adiabatic transition between the +1 and 0 states (dashed blue arrow).}
\end{figure}

In this work, we investigate how a \emph{free} current flows in 1D: as illustrated in \fref{fig:sketch}, starting from  a system initially prepared in a well-defined current state in a ring trap with a barrier, we follow the current  dynamics with the aim of elucidating the dissipation mechanisms.
Our study concerns both zero- and finite temperature gases, both at weak and strong interactions. We show that the dynamical behavior can be interpreted as a dual of the Josephson effect, occurring among angular momentum states. Depending on the barrier strength and the temperature regime we observe current oscillations, self-trapping or decay. In the weakly interacting regime, we  show that the observed dynamics correspond to self-trapping among angular momentum states at zero temperature, and that the decay of the currents at finite temperature involves dark solitons. For strong interactions, we show that  coherent quantum phase slips dominate the current dynamics at zero temperature, and incoherent ones take over at finite temperature.

\paragraph{Model}
We consider $N$ bosons of mass $m$ with repulsive contact interactions on a ring of circumference $L$ with periodic boundary conditions, i.e. the Lieb-Liniger model, generalized to include the presence of an external barrier potential $V(x)$. The Hamiltonian reads:
\begin{equation}
\hat{\mathcal{H}}=\int_0^L dx\,\hat{\Psi}^\dagger\left(-\frac{\hbar^2}{2m}\frac{\partial^2}{\partial x^2}+V(x)+\frac{g}{2}\hat{\Psi}^\dagger\hat{\Psi}\right)\hat{\Psi},
\label{eqn:fullH}
\end{equation}
where $\hat{\Psi}$ is the bosonic field operator,  $n=N/L$ the average density,   with  total number of particles $N=\int_0^L dx\,\braket{\hat{\Psi}^\dagger\hat{\Psi}}$.
This model describes e.g. ultra-cold atoms confined in a tight ring trap. In this case  $g=2\hbar \omega_\perp a_s$ is the 1D interaction strength, $\omega_\perp$  the radial confinement frequency  and $a_s$ the 3D $s$-wave scattering length. In the following we  consider either a delta potential  $V(x)=\alpha\delta(x)$, for which analytical results can be obtained, or a Gaussian potential $V(x)=V_0\exp{\left(-\frac{x^2}{2\sigma^2}\right)}$, realistic from the experimental point of view. For homogeneous 1D gases the equilibrium properties at finite temperature are captured by two dimensionless parameters~\cite{Kheruntsyan2003}: $\gamma = \frac{mg}{\hbar^2 n}$ quantifying the interaction regime from weak ($\gamma\ll1$) to strong ($\gamma\gg1$), and the reduced temperature $\tau=\frac{T}{T_d\gamma^2}$, where $T_d=\hbar^2n^2/2mk_B$ is the quantum degeneracy temperature.

\paragraph{Quench protocol}
Our goal is to study the dynamics of the particle current in the presence of a barrier. We first prepare the system in an equilibrium state $\Psi_0$ in the presence of the static barrier potential. This results in a state with no current. Specific details on the implementation depend on the interaction regime and are given later. We then quench the current by phase imprinting a specific circulation onto the many-body wavefunction: $\Psi_0(x_1,...x_N)\to\Psi_1(x_1,...x_N)=\Psi_0\times e^{i2\pi\ell \sum_j x_j/L}$. Note that this process can be implemented in experiments using specific light potentials according to various available schemes~\cite{Moulder2012,Kumar2018}.
We then monitor the current by computing the average of the current operator per particle:
\begin{equation}
J(t)=-i \frac{\hbar}{2m}\frac{1}{N}\int_0^L \frac{dx}{L}\,\Braket{\hat{\Psi}^\dagger\partial_x\hat{\Psi}-\left(\partial_x\hat{\Psi}^\dagger\right)\hat{\Psi}}.
\end{equation}
The time evolution following the quench is described by different approaches depending on the interaction and temperature regimes:  (i) at $T=0$ and for a weakly interacting gas ($\gamma\ll1$) we rely on the Gross-Pitaevskii equation (GPE) numerical solution and on an analytical two-mode model adapted from~\cite{Smerzi1997}; (ii) at $T>0$ and $\gamma\ll1$ we use the Projected Gross-Pitaevskii equation (PGPE) formalism~\cite{Davis2001,Blakie2008,Berloff2014a} and (iii) at $\gamma\gg1$ we use an  exact  time-dependent Bose-Fermi mapping describing the infinitely strong interaction Tonks-Girardeau (TG) limit for the whole temperature range~\cite{Girardeau_1960,Wright_2000,Girardeau_2005}, focusing on a quench with circulation $\ell=1$ \footnote{The calculation for $\ell=2$ yields no qualitative difference in the current dynamics, just a faster decay of the oscillations.}.

\begin{figure*}[t!]
\includegraphics[width=\linewidth]{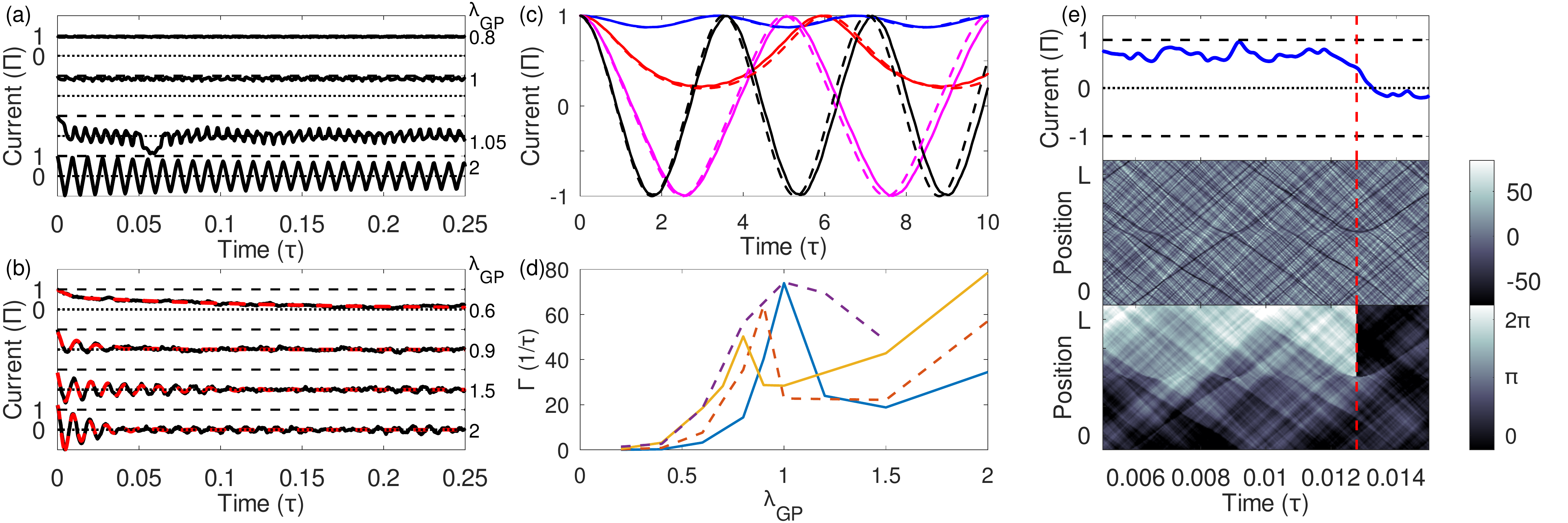}%
\caption{\label{fig:PGPE}
(color online) Classical field simulations of the  quench dynamics in the mean-field regime for $g=20\times \hbar^2/(mL)$ and $N=1000$ (corresponding to  $\gamma=0.02$). (a) Average current per particle (black solid lines, in units of $\Pi=\hbar/(Nm)$) as a function of time (in units of $\tau=mL^2/\hbar$), at $T=0$.
The horizontal black dotted (dashed) lines indicate the values  $J=0$  ($\pm1$).
From top to bottom:  $\lambda_{\rm GP}=\{0.8,1,1.05,2\}$.
(b) Current at $T=\mu_0/k_B$, averaged over 100 realizations of the classical field, for barrier strengths $\lambda_{\rm GP}=\{0.6,0.9,1.5,2\}$, black solid lines: simulations, red dashed curves:  fits from the model function  $J(t)=Ae^{-\Gamma_At}+B\cos{[\omega t+\phi]}e^{-\Gamma_Bt}$.
(c) Current for  $\gamma=2\times10^{-5}$ at $T=0$,  simulations (solid lines) and  two-mode model (dashed lines) for $\lambda_{\rm GP}=\{0.05,0.1,0.15,0.2\}$ (blue, red, magenta, black, respectively).
(d) Damping rate $\Gamma$ (in units of $1/\tau$) (extracted from the fit, maximum among $\Gamma_A$ and $\Gamma_B$)  as a function of $\lambda_{\rm GP}$ for $T=\{0.5,1,1.5,1.75\}\times\mu_0/k_B$ (solid blue, dashed red, solid yellow, dashed violet, respectively).
(e) Zoom on a single classical field trajectory, at $T=\mu_0/k_B$ and $\lambda_{\rm GP}=0.6$, evidencing a phase slip: a jump in the current (top panel) corresponds to the reflection of a slow soliton at the barrier, visible in the density deviation map \cite{DensityDev} (middle panel) and  to a singularity in the phase profile (bottom panel).
}
\end{figure*}

In the weakly interacting limit we scale the Gaussian barrier strength relative to the chemical potential, i.e. we define $\lambda_{\rm GP}=V_0/\mu_0$ with $\mu_0=gn$  the chemical potential of the homogeneous annular gas. Figure~\ref{fig:PGPE} illustrates  our simulation results in the weakly interacting regime as a function of $\lambda_{\rm GP}$ for a relatively narrow barrier of width $\sigma=L/50$, yet larger than the healing length $\xi=\hbar/\sqrt{2mgn}\simeq\sigma/4$. At zero temperature we observe in \fref{fig:PGPE}(a) that the current remains very close to the initial quenched circulating state for weak to moderate barriers, up to $\lambda_{\rm GP}\sim1$. Above this critical value, we observe a fast decay of the current, followed by oscillations around the 0 value. This is very similar to what has been obtained in 2D simulations~\cite{Mathey2014}. The new feature of the 1D mean-field regime is the emergence of current oscillations at large barriers. As we discuss here below, this behavior can be interpreted as the transition from  self-trapping to Josephson oscillations {\em of the currents}, in analogy to the well known Josephson effect for particle imbalance predicted in \cite{Smerzi1997} and  experimentally observed using ultra-cold atoms confined in a double well trap \cite{Albiez2005}.
In  essence \cite{supplemental}, we derive a fully analytical two-mode model for two current states and show that this accurately captures the Gross-Pitaevskii dynamics at zero temperature and very weak interactions (see \fref{fig:PGPE}(c)). This model predicts a transition from self-trapping to Josephson oscillations for a critical value $\lambda_{\rm GP}^c$ that depends on the interaction strength as in~\cite{Smerzi1997}. Interestingly, a two-mode model based on current states in the linear regime also accurately describes the  dynamics of vortex nucleation in stirred condensates \cite{Caradoc-Davies1999}.
Although the two-mode model breaks down for large barrier or higher (but still weak) interactions due to the spread of the mean-field wavefunction onto many single particle orbitals, we observe the same qualitative behavior in the simulations. Indeed, surprisingly, the current always oscillates regularly at large barriers (bottom curve of Fig.~\ref{fig:PGPE}(a)), with a non-sinusoidal (piecewise linear)  shape and very small damping rate. These oscillations can be understood by casting the GPE into the superfluid hydrodynamic form:   transport of matter occurs via a density fluctuation corresponding to a 
shock wave  \cite{Hakim1997}, propagating at the speed of sound on top of a moving fluid.

For temperature $T=\mu_0/k_B$, corresponding to the quasi-condensate regime~\cite{Kheruntsyan2003}, the dynamics of the current are quite different from the zero-temperature case, see \fref{fig:PGPE}(b). At low barriers, i.e. $\lambda_{\rm GP}\leq0.5$, we observe an exponential decay of the current with a decay rate increasing with the barrier strength. For larger barriers we observe damped oscillations of the current. In this regime thermal phase slips  occur deterministically at the position of the barrier, where the density vanishes. The transition from exponential to damped oscillation decay is  observed for all our temperatures in the range $0.5\leq k_BT/\mu_0 \leq 2.5$. \fref{fig:PGPE}(d) displays the value of the damping rate $\Gamma$ given by the fit \cite{Fitting} for increasing temperatures, in the range $0.5\leq k_BT/\mu_0\leq 1.75$. The damping rate increases with temperature, displaying a non-monotonous dependence on  the barrier strength, with a  maximum at the crossover between the two decay regimes. The crossover occurs at lower barrier strength for larger temperatures,  consistent with the thermal activation of solitons, as we discuss below.

In order to elucidate the mechanisms for the current decay, \fref{fig:PGPE}(e) shows  a \emph{single} classical field trajectory, showing  many spontaneous thermal gray solitons~\cite{Karpiuk2012}. While most of the solitons present a small density dip, hence are fast and  are transmitted through the barrier~\cite{Bilas2005}, we notice that the current undergoes discrete jumps each time a soliton is reflected  on the barrier: in this case, when the soliton reaches zero velocity the density profile vanishes, allowing for a phase slip to occur.
This corresponds to the adiabatic process indicated by the dashed blue line on \fref{fig:sketch}(b).
As the temperature increases, the probability to find slow solitons increases and the jumps occur more and more frequently, resulting in an increase of the decay rate, as seen in \fref{fig:PGPE}(d). Finally, as the barrier couples the soliton dynamics to the long wavelength sound excitations~\cite{Bilas2005} we expect this process to be intrinsically stochastic, thus resulting in an exponential decay of the average current as observed.


\begin{figure*}[t!]
\includegraphics[width=\linewidth]{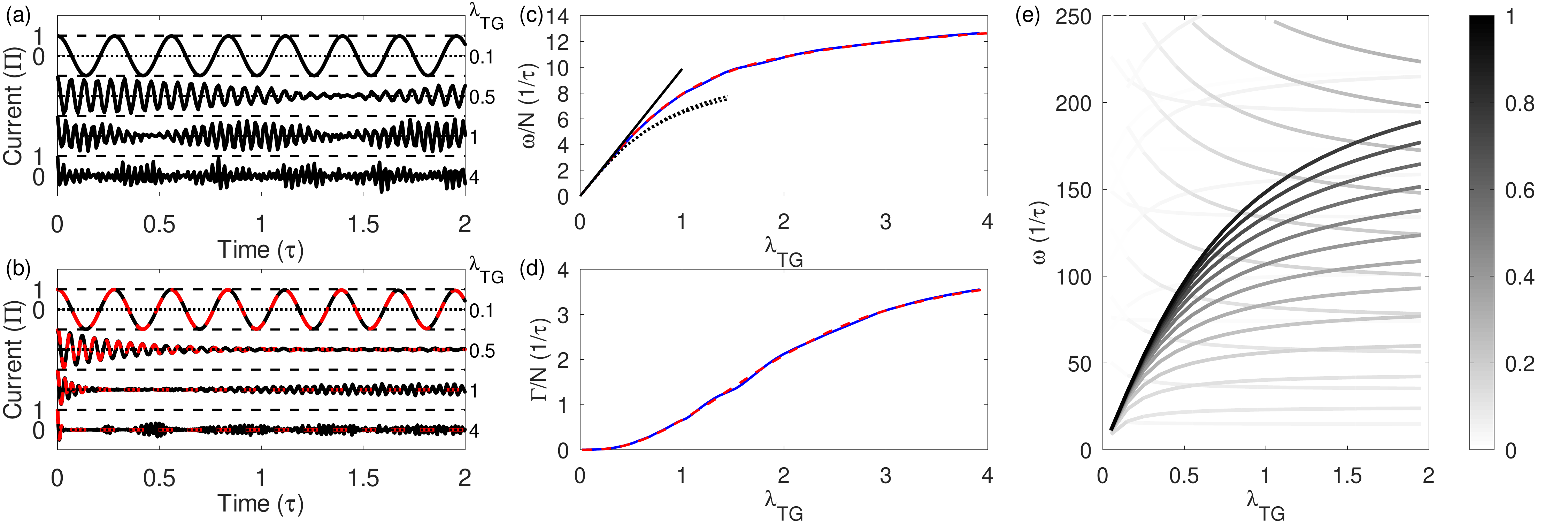}%
\caption{\label{fig:Tonks} (Color online) Exact solutions in the Tonks-Girardeau regime. (a) Average current per particle (in units of $\Pi=\hbar/Nm$) vs.~time (in units of $\tau=mL^2/\hbar$) after the quench for $N=23$, at $T=0$, for barrier strength $\lambda_{\rm TG}=\{0.1,0.5,1,4\}$.
The horizontal black dotted (dashed) lines indicate the values  $J=0$  ($\pm1$).
(b) Current at $T=E_F/k_B$ (black solid) for  $\lambda_{\rm TG}=\{0.1,0.5,1,4\}$ from top to bottom and fits (red-dashes, same fitting function as in Fig.\ref{fig:PGPE}).
(c) Frequency $\omega/N$ and (d) damping rate $\Gamma/N$ obtained from the fit vs.~$\lambda_{\rm TG}$, for $N=11$ (solid blue) and $N=23$ (dashed red). Other curves  in (c):  frequency for universal Rabi oscillations  $\omega_R=\pi^2N\lambda_{\rm TG}$ (black solid) and first excitation frequency at the Fermi sphere (black dashed) \cite{supplemental}.
(e) Frequency of the excitations  produced in the quench (relative amplitude in colormap) vs.~$\lambda_{\rm TG}$ for $N=23$ at $T=0$.
}
\end{figure*}

The description of current dynamics as dual of the Josephson effect persists at  strong interactions. In this regime, the classical picture does not apply,  rather, we  show below that the dynamics correspond to quantum coherent oscillations among angular momentum states  (see \cite{Astafiev_2012} for the analog phenomenon in superconductors). We describe the dynamics of the current in the strongly interacting limit $\gamma\gg 1$  using the  exact Tonks-Girardeau solution, which maps the interacting bosons onto a Fermi gas. In  the TG regime the relevant dimensionless barrier strength is $\lambda_{\rm TG}=V_b/E_F$, with $V_b=\alpha n$ being the barrier associated energy and $E_F=\hbar^2n^2\pi^2/2m$ being the Fermi energy, corresponding to the zero-temperature chemical potential for systems displaying fermionization \cite{supplemental}. At zero temperature, \fref{fig:Tonks}(a), we note that for weak barriers, $\lambda_{\rm TG}\ll1$, in contrast to the weakly interacting regime, there is no self-trapping, rather, the current undergoes Rabi-like oscillations. These oscillations correspond to coherent quantum phase slips due to backscattering  induced by the barrier, which breaks rotation symmetry thus coupling different angular momentum states \cite{Moij2006,Cominotti2014}. Microscopically, it
corresponds to dynamical processes involving the whole Fermi sphere, i.e. multiple-particle  hole excitations where each particle coherently undergoes oscillations of angular momentum from $L_z=\hbar$ to $L_z=-\hbar$. At increasing barrier strength, an envelope appears on top of the current oscillations, degrading the Rabi oscillations. This envelope  originates from  the population of  higher-energy modes, each transition being characterized by a different frequency (see \fref{fig:Tonks}(e) and \cite{supplemental}), leading to a mode-mode coupling and dephasing, and correspondingly more complex current oscillations.


At finite temperatures the quench dynamics of the current involve  high-energy  excitations with   amplitude  weighted by the Fermi distribution \cite{supplemental}.  The resulting dynamics correspond to an effective damping of the current oscillations with a exponential decay, see \fref{fig:Tonks}(b), corresponding  to the effect of incoherent phase slips.  The revivals observed for large  barrier  at zero temperature are highly suppressed  due to the  thermal excitations. In \fref{fig:Tonks}(d) we show the decay rate $\Gamma$ of the persistent currents as a function of the barrier strength \cite{Fitting}. We find that the decay of persistent currents grows monotonically with the barrier strength, since more and more  excitations are involved in the dynamics as the  barrier strength increases. In \fref{fig:Tonks}(c) we show the oscillation frequency as a function of $\lambda_{\rm TG}$ and observe  that at increasing barrier strength the frequency   crosses over from a Rabi-like regime with $\omega=\pi^2N \lambda_{\rm TG}$ to a Josephson-like regime with $\omega\propto \sqrt{\lambda_{\rm TG}}$, in agreement with the  predictions of  the low-energy Luttinger liquid theory \cite{Polo_2018}. Quite generally, while our results have been derived for infinite interaction strength, the predictions of the TG model, including quantum fluctuations in an exact way,  are expected to closely describe a Bose gas at strong interactions.

In conclusion,  we have shown that the dynamical evolution following a phase imprinting induces oscillations of the current in a 1D ring, associated to a rich excitation pattern, which can be  described by a dual Josephson dynamics. At weak interactions and finite temperature we observe the formation of both sound waves and of thermally activated dark solitons. We find that  phase-slippage  occurs incoherently when the solitons are reflected by  the barrier. In the strongly interacting regime at zero temperature we find coherent Rabi oscillations indicating quantum coherent phase slips, which are degraded by mode dephasing at large barrier strength or by thermal fluctuations at finite temperature. In the weakly-interacting limit we find  self-trapping of current states, while no self-trapping is found at infinitely strong interactions, where quantum fluctuations dominate. 

The dual Josephson picture is a new paradigm for dynamics of  atomtronics circuits in which a current state encodes quantum information. Our work evidences the importance of the dynamics of the current in a 1D system, which  can be accurately measured using existing experimental tools: an  interferometric measurement  accessing  the local currents~\cite{Corman2014,Eckel2014a} or long wavelength excitations~\cite{Marti2014,Kumar2016}. The stochastic decay of the current in 1D via phase slips is  reminiscent of the stochastic decay due to vortex/anti-vortex recombination in 2D or 3D systems \cite{Piazza2009}, where, however,  oscillations are  strongly damped by vortex creation \cite{Mathey2014}.
The  main difference between  1D and the higher-dimensional counterparts is that in the former case  the current dynamics are more robust:  at weak interactions, the solitons properties are gradually degraded by the several interactions with the barrier, mainly by sound wave radiation~\cite{Bilas2005}, and at strong interactions we observe the coherent dynamics of all particles. In outlook, it would be very interesting to investigate how the self-trapping disappears for large but finite interactions as well as  to study the crossover to a quasi-1D geometry to explore the role of radial modes in the decay  dynamics.

\begin{acknowledgments}
We thank Maxim Olshanii and Jook Walraven for stimulating discussions.
We acknowledge financial support from the ANR project SuperRing (Grant
No.  ANR-15-CE30-0012). LPL is a member DIM SIRTEQ 
(Science et Ing\'enierie en R\'egion \^Ile-de-France pour les Technologies 
Quantiques).
\end{acknowledgments}

\bibliography{biblio.bib}

\begin{thebibliography}{66}%
\makeatletter
\providecommand \@ifxundefined [1]{%
 \@ifx{#1\undefined}
}%
\providecommand \@ifnum [1]{%
 \ifnum #1\expandafter \@firstoftwo
 \else \expandafter \@secondoftwo
 \fi
}%
\providecommand \@ifx [1]{%
 \ifx #1\expandafter \@firstoftwo
 \else \expandafter \@secondoftwo
 \fi
}%
\providecommand \natexlab [1]{#1}%
\providecommand \enquote  [1]{``#1''}%
\providecommand \bibnamefont  [1]{#1}%
\providecommand \bibfnamefont [1]{#1}%
\providecommand \citenamefont [1]{#1}%
\providecommand \href@noop [0]{\@secondoftwo}%
\providecommand \href [0]{\begingroup \@sanitize@url \@href}%
\providecommand \@href[1]{\@@startlink{#1}\@@href}%
\providecommand \@@href[1]{\endgroup#1\@@endlink}%
\providecommand \@sanitize@url [0]{\catcode `\\12\catcode `\$12\catcode
  `\&12\catcode `\#12\catcode `\^12\catcode `\_12\catcode `\%12\relax}%
\providecommand \@@startlink[1]{}%
\providecommand \@@endlink[0]{}%
\providecommand \url  [0]{\begingroup\@sanitize@url \@url }%
\providecommand \@url [1]{\endgroup\@href {#1}{\urlprefix }}%
\providecommand \urlprefix  [0]{URL }%
\providecommand \Eprint [0]{\href }%
\providecommand \doibase [0]{http://dx.doi.org/}%
\providecommand \selectlanguage [0]{\@gobble}%
\providecommand \bibinfo  [0]{\@secondoftwo}%
\providecommand \bibfield  [0]{\@secondoftwo}%
\providecommand \translation [1]{[#1]}%
\providecommand \BibitemOpen [0]{}%
\providecommand \bibitemStop [0]{}%
\providecommand \bibitemNoStop [0]{.\EOS\space}%
\providecommand \EOS [0]{\spacefactor3000\relax}%
\providecommand \BibitemShut  [1]{\csname bibitem#1\endcsname}%
\let\auto@bib@innerbib\@empty
\bibitem [{\citenamefont {Ramanathan}\ \emph {et~al.}(2011)\citenamefont
  {Ramanathan}, \citenamefont {Wright}, \citenamefont {Muniz}, \citenamefont
  {Zelan}, \citenamefont {Hill}, \citenamefont {Lobb}, \citenamefont
  {Helmerson}, \citenamefont {Phillips},\ and\ \citenamefont
  {Campbell}}]{Ramanathan2011}%
  \BibitemOpen
  \bibfield  {author} {\bibinfo {author} {\bibfnamefont {A.}~\bibnamefont
  {Ramanathan}}, \bibinfo {author} {\bibfnamefont {K.~C.}\ \bibnamefont
  {Wright}}, \bibinfo {author} {\bibfnamefont {S.~R.}\ \bibnamefont {Muniz}},
  \bibinfo {author} {\bibfnamefont {M.}~\bibnamefont {Zelan}}, \bibinfo
  {author} {\bibfnamefont {W.~T.}\ \bibnamefont {Hill}}, \bibinfo {author}
  {\bibfnamefont {C.~J.}\ \bibnamefont {Lobb}}, \bibinfo {author}
  {\bibfnamefont {K.}~\bibnamefont {Helmerson}}, \bibinfo {author}
  {\bibfnamefont {W.~D.}\ \bibnamefont {Phillips}}, \ and\ \bibinfo {author}
  {\bibfnamefont {G.~K.}\ \bibnamefont {Campbell}},\ }\href {\doibase
  10.1103/PhysRevLett.106.130401} {\bibfield  {journal} {\bibinfo  {journal}
  {Phys. Rev. Lett.}\ }\textbf {\bibinfo {volume} {106}},\ \bibinfo {pages}
  {130401} (\bibinfo {year} {2011})}\BibitemShut {NoStop}%
\bibitem [{\citenamefont {Moulder}\ \emph {et~al.}(2012)\citenamefont
  {Moulder}, \citenamefont {Beattie}, \citenamefont {Smith}, \citenamefont
  {Tammuz},\ and\ \citenamefont {Hadzibabic}}]{Moulder2012}%
  \BibitemOpen
  \bibfield  {author} {\bibinfo {author} {\bibfnamefont {S.}~\bibnamefont
  {Moulder}}, \bibinfo {author} {\bibfnamefont {S.}~\bibnamefont {Beattie}},
  \bibinfo {author} {\bibfnamefont {R.~P.}\ \bibnamefont {Smith}}, \bibinfo
  {author} {\bibfnamefont {N.}~\bibnamefont {Tammuz}}, \ and\ \bibinfo {author}
  {\bibfnamefont {Z.}~\bibnamefont {Hadzibabic}},\ }\href {\doibase
  10.1103/PhysRevA.86.013629} {\bibfield  {journal} {\bibinfo  {journal} {Phys.
  Rev. A}\ }\textbf {\bibinfo {volume} {86}},\ \bibinfo {pages} {013629}
  (\bibinfo {year} {2012})}\BibitemShut {NoStop}%
\bibitem [{\citenamefont {{Husmann}}\ \emph {et~al.}(2015)\citenamefont
  {{Husmann}}, \citenamefont {{Uchino}}, \citenamefont {{Krinner}},
  \citenamefont {{Lebrat}}, \citenamefont {{Giamarchi}}, \citenamefont
  {{Esslinger}},\ and\ \citenamefont {{Brantut}}}]{Husmann_2015}%
  \BibitemOpen
  \bibfield  {author} {\bibinfo {author} {\bibfnamefont {D.}~\bibnamefont
  {{Husmann}}}, \bibinfo {author} {\bibfnamefont {S.}~\bibnamefont {{Uchino}}},
  \bibinfo {author} {\bibfnamefont {S.}~\bibnamefont {{Krinner}}}, \bibinfo
  {author} {\bibfnamefont {M.}~\bibnamefont {{Lebrat}}}, \bibinfo {author}
  {\bibfnamefont {T.}~\bibnamefont {{Giamarchi}}}, \bibinfo {author}
  {\bibfnamefont {T.}~\bibnamefont {{Esslinger}}}, \ and\ \bibinfo {author}
  {\bibfnamefont {J.-P.}\ \bibnamefont {{Brantut}}},\ }\href {\doibase
  10.1126/science.aac9584} {\bibfield  {journal} {\bibinfo  {journal}
  {Science}\ }\textbf {\bibinfo {volume} {350}},\ \bibinfo {pages} {1498}
  (\bibinfo {year} {2015})}\BibitemShut {NoStop}%
\bibitem [{\citenamefont {Mathey}\ and\ \citenamefont
  {Mathey}(2016)}]{Mathey_2016}%
  \BibitemOpen
  \bibfield  {author} {\bibinfo {author} {\bibfnamefont {A.~C.}\ \bibnamefont
  {Mathey}}\ and\ \bibinfo {author} {\bibfnamefont {L.}~\bibnamefont
  {Mathey}},\ }\href {\doibase 10.1088/1367-2630/18/5/055016} {\bibfield
  {journal} {\bibinfo  {journal} {New Journal of Physics}\ }\textbf {\bibinfo
  {volume} {18}},\ \bibinfo {pages} {055016} (\bibinfo {year}
  {2016})}\BibitemShut {NoStop}%
\bibitem [{\citenamefont {Seaman}\ \emph {et~al.}(2007)\citenamefont {Seaman},
  \citenamefont {Kr\"amer}, \citenamefont {Anderson},\ and\ \citenamefont
  {Holland}}]{Seaman_2007}%
  \BibitemOpen
  \bibfield  {author} {\bibinfo {author} {\bibfnamefont {B.~T.}\ \bibnamefont
  {Seaman}}, \bibinfo {author} {\bibfnamefont {M.}~\bibnamefont {Kr\"amer}},
  \bibinfo {author} {\bibfnamefont {D.~Z.}\ \bibnamefont {Anderson}}, \ and\
  \bibinfo {author} {\bibfnamefont {M.~J.}\ \bibnamefont {Holland}},\ }\href
  {\doibase 10.1103/PhysRevA.75.023615} {\bibfield  {journal} {\bibinfo
  {journal} {Phys. Rev. A}\ }\textbf {\bibinfo {volume} {75}},\ \bibinfo
  {pages} {023615} (\bibinfo {year} {2007})}\BibitemShut {NoStop}%
\bibitem [{\citenamefont {Amico}\ \emph {et~al.}(2014)\citenamefont {Amico},
  \citenamefont {Aghamalyan}, \citenamefont {Auksztol}, \citenamefont {Crepaz},
  \citenamefont {Dumke},\ and\ \citenamefont {Kwek}}]{Amico2014}%
  \BibitemOpen
  \bibfield  {author} {\bibinfo {author} {\bibfnamefont {L.}~\bibnamefont
  {Amico}}, \bibinfo {author} {\bibfnamefont {D.}~\bibnamefont {Aghamalyan}},
  \bibinfo {author} {\bibfnamefont {F.}~\bibnamefont {Auksztol}}, \bibinfo
  {author} {\bibfnamefont {H.}~\bibnamefont {Crepaz}}, \bibinfo {author}
  {\bibfnamefont {R.}~\bibnamefont {Dumke}}, \ and\ \bibinfo {author}
  {\bibfnamefont {L.~C.}\ \bibnamefont {Kwek}},\ }\href {\doibase
  10.1038/srep04298} {\bibfield  {journal} {\bibinfo  {journal} {Scientific
  reports}\ }\textbf {\bibinfo {volume} {4}},\ \bibinfo {pages} {4298}
  (\bibinfo {year} {2014})}\BibitemShut {NoStop}%
\bibitem [{\citenamefont {Amico}\ \emph {et~al.}(2017)\citenamefont {Amico},
  \citenamefont {Birkl}, \citenamefont {Boshier},\ and\ \citenamefont
  {Kwek}}]{Amico_2017}%
  \BibitemOpen
  \bibfield  {author} {\bibinfo {author} {\bibfnamefont {L.}~\bibnamefont
  {Amico}}, \bibinfo {author} {\bibfnamefont {G.}~\bibnamefont {Birkl}},
  \bibinfo {author} {\bibfnamefont {M.}~\bibnamefont {Boshier}}, \ and\
  \bibinfo {author} {\bibfnamefont {L.-C.}\ \bibnamefont {Kwek}},\ }\href
  {\doibase 10.1088/1367-2630/aa5a6d} {\bibfield  {journal} {\bibinfo
  {journal} {New Journal of Physics}\ }\textbf {\bibinfo {volume} {19}},\
  \bibinfo {pages} {020201} (\bibinfo {year} {2017})}\BibitemShut {NoStop}%
\bibitem [{\citenamefont {Gauthier}\ \emph {et~al.}(2019)\citenamefont
  {Gauthier}, \citenamefont {Szigeti}, \citenamefont {Reeves}, \citenamefont
  {Baker}, \citenamefont {Bell}, \citenamefont {Rubinsztein-Dunlop},
  \citenamefont {Davis},\ and\ \citenamefont {Neely}}]{Gauthier2019}%
  \BibitemOpen
  \bibfield  {author} {\bibinfo {author} {\bibfnamefont {G.}~\bibnamefont
  {Gauthier}}, \bibinfo {author} {\bibfnamefont {S.~S.}\ \bibnamefont
  {Szigeti}}, \bibinfo {author} {\bibfnamefont {M.~T.}\ \bibnamefont {Reeves}},
  \bibinfo {author} {\bibfnamefont {M.}~\bibnamefont {Baker}}, \bibinfo
  {author} {\bibfnamefont {T.~A.}\ \bibnamefont {Bell}}, \bibinfo {author}
  {\bibfnamefont {H.}~\bibnamefont {Rubinsztein-Dunlop}}, \bibinfo {author}
  {\bibfnamefont {M.~J.}\ \bibnamefont {Davis}}, \ and\ \bibinfo {author}
  {\bibfnamefont {T.~W.}\ \bibnamefont {Neely}},\ }\href
  {http://arxiv.org/abs/1903.04086} {\ ,\ \bibinfo {pages} {1} (\bibinfo {year}
  {2019})},\ \Eprint {http://arxiv.org/abs/1903.04086} {arXiv:1903.04086}
  \BibitemShut {NoStop}%
\bibitem [{\citenamefont {Madison}\ \emph {et~al.}(2001)\citenamefont
  {Madison}, \citenamefont {Chevy}, \citenamefont {Bretin},\ and\ \citenamefont
  {Dalibard}}]{Madison2001}%
  \BibitemOpen
  \bibfield  {author} {\bibinfo {author} {\bibfnamefont {K.~W.}\ \bibnamefont
  {Madison}}, \bibinfo {author} {\bibfnamefont {F.}~\bibnamefont {Chevy}},
  \bibinfo {author} {\bibfnamefont {V.}~\bibnamefont {Bretin}}, \ and\ \bibinfo
  {author} {\bibfnamefont {J.}~\bibnamefont {Dalibard}},\ }\href {\doibase
  10.1103/PhysRevLett.86.4443} {\bibfield  {journal} {\bibinfo  {journal}
  {Phys. Rev. Lett.}\ }\textbf {\bibinfo {volume} {86}},\ \bibinfo {pages}
  {4443} (\bibinfo {year} {2001})}\BibitemShut {NoStop}%
\bibitem [{\citenamefont {Penckwitt}\ \emph {et~al.}(2002)\citenamefont
  {Penckwitt}, \citenamefont {Ballagh},\ and\ \citenamefont
  {Gardiner}}]{Penckwitt2002}%
  \BibitemOpen
  \bibfield  {author} {\bibinfo {author} {\bibfnamefont {A.~A.}\ \bibnamefont
  {Penckwitt}}, \bibinfo {author} {\bibfnamefont {R.~J.}\ \bibnamefont
  {Ballagh}}, \ and\ \bibinfo {author} {\bibfnamefont {C.~W.}\ \bibnamefont
  {Gardiner}},\ }\href {\doibase 10.1103/PhysRevLett.89.260402} {\bibfield
  {journal} {\bibinfo  {journal} {Phys. Rev. Lett.}\ }\textbf {\bibinfo
  {volume} {89}},\ \bibinfo {pages} {260402} (\bibinfo {year}
  {2002})}\BibitemShut {NoStop}%
\bibitem [{\citenamefont {Lobo}\ \emph {et~al.}(2004)\citenamefont {Lobo},
  \citenamefont {Sinatra},\ and\ \citenamefont {Castin}}]{Lobo2004}%
  \BibitemOpen
  \bibfield  {author} {\bibinfo {author} {\bibfnamefont {C.}~\bibnamefont
  {Lobo}}, \bibinfo {author} {\bibfnamefont {A.}~\bibnamefont {Sinatra}}, \
  and\ \bibinfo {author} {\bibfnamefont {Y.}~\bibnamefont {Castin}},\ }\href
  {\doibase 10.1103/PhysRevLett.92.020403} {\bibfield  {journal} {\bibinfo
  {journal} {Physical review letters}\ }\textbf {\bibinfo {volume} {92}},\
  \bibinfo {pages} {020403} (\bibinfo {year} {2004})}\BibitemShut {NoStop}%
\bibitem [{\citenamefont {Dubessy}\ \emph {et~al.}(2012)\citenamefont
  {Dubessy}, \citenamefont {Liennard}, \citenamefont {Pedri},\ and\
  \citenamefont {Perrin}}]{Dubessy2012}%
  \BibitemOpen
  \bibfield  {author} {\bibinfo {author} {\bibfnamefont {R.}~\bibnamefont
  {Dubessy}}, \bibinfo {author} {\bibfnamefont {T.}~\bibnamefont {Liennard}},
  \bibinfo {author} {\bibfnamefont {P.}~\bibnamefont {Pedri}}, \ and\ \bibinfo
  {author} {\bibfnamefont {H.}~\bibnamefont {Perrin}},\ }\href {\doibase
  10.1103/PhysRevA.86.011602} {\bibfield  {journal} {\bibinfo  {journal} {Phys.
  Rev. A}\ }\textbf {\bibinfo {volume} {86}},\ \bibinfo {pages} {011602(R)}
  (\bibinfo {year} {2012})}\BibitemShut {NoStop}%
\bibitem [{\citenamefont {Wright}\ \emph
  {et~al.}(2013{\natexlab{a}})\citenamefont {Wright}, \citenamefont
  {Blakestad}, \citenamefont {Lobb}, \citenamefont {Phillips},\ and\
  \citenamefont {Campbell}}]{Wright2013a}%
  \BibitemOpen
  \bibfield  {author} {\bibinfo {author} {\bibfnamefont {K.~C.}\ \bibnamefont
  {Wright}}, \bibinfo {author} {\bibfnamefont {R.~B.}\ \bibnamefont
  {Blakestad}}, \bibinfo {author} {\bibfnamefont {C.~J.}\ \bibnamefont {Lobb}},
  \bibinfo {author} {\bibfnamefont {W.~D.}\ \bibnamefont {Phillips}}, \ and\
  \bibinfo {author} {\bibfnamefont {G.~K.}\ \bibnamefont {Campbell}},\ }\href
  {\doibase 10.1103/PhysRevA.88.063633} {\bibfield  {journal} {\bibinfo
  {journal} {Phys. Rev. A}\ }\textbf {\bibinfo {volume} {88}},\ \bibinfo
  {pages} {063633} (\bibinfo {year} {2013}{\natexlab{a}})}\BibitemShut
  {NoStop}%
\bibitem [{\citenamefont {Wright}\ \emph
  {et~al.}(2013{\natexlab{b}})\citenamefont {Wright}, \citenamefont
  {Blakestad}, \citenamefont {Lobb}, \citenamefont {Phillips},\ and\
  \citenamefont {Campbell}}]{Wright2013}%
  \BibitemOpen
  \bibfield  {author} {\bibinfo {author} {\bibfnamefont {K.~C.}\ \bibnamefont
  {Wright}}, \bibinfo {author} {\bibfnamefont {R.~B.}\ \bibnamefont
  {Blakestad}}, \bibinfo {author} {\bibfnamefont {C.~J.}\ \bibnamefont {Lobb}},
  \bibinfo {author} {\bibfnamefont {W.~D.}\ \bibnamefont {Phillips}}, \ and\
  \bibinfo {author} {\bibfnamefont {G.~K.}\ \bibnamefont {Campbell}},\ }\href
  {\doibase 10.1103/PhysRevLett.110.025302} {\bibfield  {journal} {\bibinfo
  {journal} {Phys. Rev. Lett.}\ }\textbf {\bibinfo {volume} {110}},\ \bibinfo
  {pages} {025302} (\bibinfo {year} {2013}{\natexlab{b}})}\BibitemShut
  {NoStop}%
\bibitem [{\citenamefont {Eckel}\ \emph
  {et~al.}(2014{\natexlab{a}})\citenamefont {Eckel}, \citenamefont {Lee},
  \citenamefont {Jendrzejewski}, \citenamefont {Murray}, \citenamefont {Clark},
  \citenamefont {Lobb}, \citenamefont {Phillips}, \citenamefont {Edwards},\
  and\ \citenamefont {Campbell}}]{Eckel2014}%
  \BibitemOpen
  \bibfield  {author} {\bibinfo {author} {\bibfnamefont {S.}~\bibnamefont
  {Eckel}}, \bibinfo {author} {\bibfnamefont {J.~G.}\ \bibnamefont {Lee}},
  \bibinfo {author} {\bibfnamefont {F.}~\bibnamefont {Jendrzejewski}}, \bibinfo
  {author} {\bibfnamefont {N.}~\bibnamefont {Murray}}, \bibinfo {author}
  {\bibfnamefont {C.~W.}\ \bibnamefont {Clark}}, \bibinfo {author}
  {\bibfnamefont {C.~J.}\ \bibnamefont {Lobb}}, \bibinfo {author}
  {\bibfnamefont {W.~D.}\ \bibnamefont {Phillips}}, \bibinfo {author}
  {\bibfnamefont {M.}~\bibnamefont {Edwards}}, \ and\ \bibinfo {author}
  {\bibfnamefont {G.~K.}\ \bibnamefont {Campbell}},\ }\href {\doibase
  10.1038/nature12958} {\bibfield  {journal} {\bibinfo  {journal} {Nature}\
  }\textbf {\bibinfo {volume} {506}},\ \bibinfo {pages} {200} (\bibinfo {year}
  {2014}{\natexlab{a}})}\BibitemShut {NoStop}%
\bibitem [{\citenamefont {Yakimenko}\ \emph {et~al.}(2015)\citenamefont
  {Yakimenko}, \citenamefont {Isaieva}, \citenamefont {Vilchinskii},\ and\
  \citenamefont {Ostrovskaya}}]{Yakimenko_2015}%
  \BibitemOpen
  \bibfield  {author} {\bibinfo {author} {\bibfnamefont {A.~I.}\ \bibnamefont
  {Yakimenko}}, \bibinfo {author} {\bibfnamefont {K.~O.}\ \bibnamefont
  {Isaieva}}, \bibinfo {author} {\bibfnamefont {S.~I.}\ \bibnamefont
  {Vilchinskii}}, \ and\ \bibinfo {author} {\bibfnamefont {E.~A.}\ \bibnamefont
  {Ostrovskaya}},\ }\href {\doibase 10.1103/PhysRevA.91.023607} {\bibfield
  {journal} {\bibinfo  {journal} {Phys. Rev. A}\ }\textbf {\bibinfo {volume}
  {91}},\ \bibinfo {pages} {023607} (\bibinfo {year} {2015})}\BibitemShut
  {NoStop}%
\bibitem [{\citenamefont {Mu\~noz Mateo}\ \emph {et~al.}(2015)\citenamefont
  {Mu\~noz Mateo}, \citenamefont {Gallem\'{\i}}, \citenamefont {Guilleumas},\
  and\ \citenamefont {Mayol}}]{Munoz2015}%
  \BibitemOpen
  \bibfield  {author} {\bibinfo {author} {\bibfnamefont {A.}~\bibnamefont
  {Mu\~noz Mateo}}, \bibinfo {author} {\bibfnamefont {A.}~\bibnamefont
  {Gallem\'{\i}}}, \bibinfo {author} {\bibfnamefont {M.}~\bibnamefont
  {Guilleumas}}, \ and\ \bibinfo {author} {\bibfnamefont {R.}~\bibnamefont
  {Mayol}},\ }\href {\doibase 10.1103/PhysRevA.91.063625} {\bibfield  {journal}
  {\bibinfo  {journal} {Phys. Rev. A}\ }\textbf {\bibinfo {volume} {91}},\
  \bibinfo {pages} {063625} (\bibinfo {year} {2015})}\BibitemShut {NoStop}%
\bibitem [{\citenamefont {Kumar}\ \emph {et~al.}(2017)\citenamefont {Kumar},
  \citenamefont {Eckel}, \citenamefont {Jendrzejewski},\ and\ \citenamefont
  {Campbell}}]{Kumar2017a}%
  \BibitemOpen
  \bibfield  {author} {\bibinfo {author} {\bibfnamefont {A.}~\bibnamefont
  {Kumar}}, \bibinfo {author} {\bibfnamefont {S.}~\bibnamefont {Eckel}},
  \bibinfo {author} {\bibfnamefont {F.}~\bibnamefont {Jendrzejewski}}, \ and\
  \bibinfo {author} {\bibfnamefont {G.~K.}\ \bibnamefont {Campbell}},\ }\href
  {\doibase 10.1103/PhysRevA.95.021602} {\bibfield  {journal} {\bibinfo
  {journal} {Phys. Rev. A}\ }\textbf {\bibinfo {volume} {95}},\ \bibinfo
  {pages} {021602(R)} (\bibinfo {year} {2017})}\BibitemShut {NoStop}%
\bibitem [{\citenamefont {Piazza}\ \emph {et~al.}(2009)\citenamefont {Piazza},
  \citenamefont {Collins},\ and\ \citenamefont {Smerzi}}]{Piazza2009}%
  \BibitemOpen
  \bibfield  {author} {\bibinfo {author} {\bibfnamefont {F.}~\bibnamefont
  {Piazza}}, \bibinfo {author} {\bibfnamefont {L.~A.}\ \bibnamefont {Collins}},
  \ and\ \bibinfo {author} {\bibfnamefont {A.}~\bibnamefont {Smerzi}},\ }\href
  {\doibase 10.1103/PhysRevA.80.021601} {\bibfield  {journal} {\bibinfo
  {journal} {Phys. Rev. A}\ }\textbf {\bibinfo {volume} {80}},\ \bibinfo
  {pages} {021601(R)} (\bibinfo {year} {2009})}\BibitemShut {NoStop}%
\bibitem [{\citenamefont {Mathey}\ \emph {et~al.}(2014)\citenamefont {Mathey},
  \citenamefont {Clark},\ and\ \citenamefont {Mathey}}]{Mathey2014}%
  \BibitemOpen
  \bibfield  {author} {\bibinfo {author} {\bibfnamefont {A.~C.}\ \bibnamefont
  {Mathey}}, \bibinfo {author} {\bibfnamefont {C.~W.}\ \bibnamefont {Clark}}, \
  and\ \bibinfo {author} {\bibfnamefont {L.}~\bibnamefont {Mathey}},\ }\href
  {\doibase 10.1103/PhysRevA.90.023604} {\bibfield  {journal} {\bibinfo
  {journal} {Phys. Rev. A}\ }\textbf {\bibinfo {volume} {90}},\ \bibinfo
  {pages} {023604} (\bibinfo {year} {2014})}\BibitemShut {NoStop}%
\bibitem [{\citenamefont {Kunimi}\ and\ \citenamefont
  {Danshita}(2017)}]{Kunimi2017}%
  \BibitemOpen
  \bibfield  {author} {\bibinfo {author} {\bibfnamefont {M.}~\bibnamefont
  {Kunimi}}\ and\ \bibinfo {author} {\bibfnamefont {I.}~\bibnamefont
  {Danshita}},\ }\href {\doibase 10.1103/PhysRevA.95.033637} {\bibfield
  {journal} {\bibinfo  {journal} {Phys. Rev. A}\ }\textbf {\bibinfo {volume}
  {95}},\ \bibinfo {pages} {033637} (\bibinfo {year} {2017})}\BibitemShut
  {NoStop}%
\bibitem [{\citenamefont {Burchianti}\ \emph {et~al.}(2018)\citenamefont
  {Burchianti}, \citenamefont {Scazza}, \citenamefont {Amico}, \citenamefont
  {Valtolina}, \citenamefont {Seman}, \citenamefont {Fort}, \citenamefont
  {Zaccanti}, \citenamefont {Inguscio},\ and\ \citenamefont
  {Roati}}]{Burchianti2018}%
  \BibitemOpen
  \bibfield  {author} {\bibinfo {author} {\bibfnamefont {A.}~\bibnamefont
  {Burchianti}}, \bibinfo {author} {\bibfnamefont {F.}~\bibnamefont {Scazza}},
  \bibinfo {author} {\bibfnamefont {A.}~\bibnamefont {Amico}}, \bibinfo
  {author} {\bibfnamefont {G.}~\bibnamefont {Valtolina}}, \bibinfo {author}
  {\bibfnamefont {J.~A.}\ \bibnamefont {Seman}}, \bibinfo {author}
  {\bibfnamefont {C.}~\bibnamefont {Fort}}, \bibinfo {author} {\bibfnamefont
  {M.}~\bibnamefont {Zaccanti}}, \bibinfo {author} {\bibfnamefont
  {M.}~\bibnamefont {Inguscio}}, \ and\ \bibinfo {author} {\bibfnamefont
  {G.}~\bibnamefont {Roati}},\ }\href {\doibase 10.1103/PhysRevLett.120.025302}
  {\bibfield  {journal} {\bibinfo  {journal} {Phys. Rev. Lett.}\ }\textbf
  {\bibinfo {volume} {120}},\ \bibinfo {pages} {025302} (\bibinfo {year}
  {2018})}\BibitemShut {NoStop}%
\bibitem [{\citenamefont {Xhani}\ \emph {et~al.}(2019)\citenamefont {Xhani},
  \citenamefont {Neri}, \citenamefont {Galantucci}, \citenamefont {Scazza},
  \citenamefont {Burchianti}, \citenamefont {Lee}, \citenamefont {Barenghi},
  \citenamefont {Trombettoni}, \citenamefont {Inguscio}, \citenamefont
  {Zaccanti}, \citenamefont {Roati},\ and\ \citenamefont
  {Proukakis}}]{Xhani2019}%
  \BibitemOpen
  \bibfield  {author} {\bibinfo {author} {\bibfnamefont {K.}~\bibnamefont
  {Xhani}}, \bibinfo {author} {\bibfnamefont {E.}~\bibnamefont {Neri}},
  \bibinfo {author} {\bibfnamefont {L.}~\bibnamefont {Galantucci}}, \bibinfo
  {author} {\bibfnamefont {F.}~\bibnamefont {Scazza}}, \bibinfo {author}
  {\bibfnamefont {A.}~\bibnamefont {Burchianti}}, \bibinfo {author}
  {\bibfnamefont {K.~L.}\ \bibnamefont {Lee}}, \bibinfo {author} {\bibfnamefont
  {C.~F.}\ \bibnamefont {Barenghi}}, \bibinfo {author} {\bibfnamefont
  {A.}~\bibnamefont {Trombettoni}}, \bibinfo {author} {\bibfnamefont
  {M.}~\bibnamefont {Inguscio}}, \bibinfo {author} {\bibfnamefont
  {M.}~\bibnamefont {Zaccanti}}, \bibinfo {author} {\bibfnamefont
  {G.}~\bibnamefont {Roati}}, \ and\ \bibinfo {author} {\bibfnamefont {N.~P.}\
  \bibnamefont {Proukakis}},\ }\href {http://arxiv.org/abs/1905.08893}
  {\bibfield  {journal} {\bibinfo  {journal} {arXiv e-prints}\ } (\bibinfo
  {year} {2019})},\ \Eprint {http://arxiv.org/abs/1905.08893}
  {arXiv:1905.08893} \BibitemShut {NoStop}%
\bibitem [{\citenamefont {Danshita}\ and\ \citenamefont
  {Polkovnikov}(2012)}]{Danshita2012a}%
  \BibitemOpen
  \bibfield  {author} {\bibinfo {author} {\bibfnamefont {I.}~\bibnamefont
  {Danshita}}\ and\ \bibinfo {author} {\bibfnamefont {A.}~\bibnamefont
  {Polkovnikov}},\ }\href {\doibase 10.1103/PhysRevA.85.023638} {\bibfield
  {journal} {\bibinfo  {journal} {Phys. Rev. A}\ }\textbf {\bibinfo {volume}
  {85}},\ \bibinfo {pages} {023638} (\bibinfo {year} {2012})}\BibitemShut
  {NoStop}%
\bibitem [{\citenamefont {D'Errico}\ \emph {et~al.}(2017)\citenamefont
  {D'Errico}, \citenamefont {Abbate},\ and\ \citenamefont
  {Modugno}}]{DErrico2017}%
  \BibitemOpen
  \bibfield  {author} {\bibinfo {author} {\bibfnamefont {C.}~\bibnamefont
  {D'Errico}}, \bibinfo {author} {\bibfnamefont {S.~S.}\ \bibnamefont
  {Abbate}}, \ and\ \bibinfo {author} {\bibfnamefont {G.}~\bibnamefont
  {Modugno}},\ }\href {\doibase 10.1098/rsta.2016.0425} {\bibfield  {journal}
  {\bibinfo  {journal} {Philosophical Transactions of the Royal Society A:
  Mathematical, Physical and Engineering Sciences}\ }\textbf {\bibinfo {volume}
  {375}},\ \bibinfo {pages} {20160425} (\bibinfo {year} {2017})}\BibitemShut
  {NoStop}%
\bibitem [{\citenamefont {Tanzi}\ \emph {et~al.}(2016)\citenamefont {Tanzi},
  \citenamefont {{Scaffidi Abbate}}, \citenamefont {Cataldini}, \citenamefont
  {Gori}, \citenamefont {Lucioni}, \citenamefont {Inguscio}, \citenamefont
  {Modugno},\ and\ \citenamefont {D'Errico}}]{Tanzi2016}%
  \BibitemOpen
  \bibfield  {author} {\bibinfo {author} {\bibfnamefont {L.}~\bibnamefont
  {Tanzi}}, \bibinfo {author} {\bibfnamefont {S.}~\bibnamefont {{Scaffidi
  Abbate}}}, \bibinfo {author} {\bibfnamefont {F.}~\bibnamefont {Cataldini}},
  \bibinfo {author} {\bibfnamefont {L.}~\bibnamefont {Gori}}, \bibinfo {author}
  {\bibfnamefont {E.}~\bibnamefont {Lucioni}}, \bibinfo {author} {\bibfnamefont
  {M.}~\bibnamefont {Inguscio}}, \bibinfo {author} {\bibfnamefont
  {G.}~\bibnamefont {Modugno}}, \ and\ \bibinfo {author} {\bibfnamefont
  {C.}~\bibnamefont {D'Errico}},\ }\href {\doibase 10.1038/srep25965}
  {\bibfield  {journal} {\bibinfo  {journal} {Scientific Reports}\ }\textbf
  {\bibinfo {volume} {6}},\ \bibinfo {pages} {25965} (\bibinfo {year}
  {2016})}\BibitemShut {NoStop}%
\bibitem [{\citenamefont {Cherny}\ \emph {et~al.}(2009)\citenamefont {Cherny},
  \citenamefont {Caux},\ and\ \citenamefont {Brand}}]{Cherny2009}%
  \BibitemOpen
  \bibfield  {author} {\bibinfo {author} {\bibfnamefont {A.~Y.}\ \bibnamefont
  {Cherny}}, \bibinfo {author} {\bibfnamefont {J.-S.}\ \bibnamefont {Caux}}, \
  and\ \bibinfo {author} {\bibfnamefont {J.}~\bibnamefont {Brand}},\ }\href
  {\doibase 10.1103/PhysRevA.80.043604} {\bibfield  {journal} {\bibinfo
  {journal} {Phys. Rev. A}\ }\textbf {\bibinfo {volume} {80}},\ \bibinfo
  {pages} {043604} (\bibinfo {year} {2009})}\BibitemShut {NoStop}%
\bibitem [{\citenamefont {Carr}\ \emph {et~al.}(2000)\citenamefont {Carr},
  \citenamefont {Clark},\ and\ \citenamefont {Reinhardt}}]{Carr2000a}%
  \BibitemOpen
  \bibfield  {author} {\bibinfo {author} {\bibfnamefont {L.~D.}\ \bibnamefont
  {Carr}}, \bibinfo {author} {\bibfnamefont {C.~W.}\ \bibnamefont {Clark}}, \
  and\ \bibinfo {author} {\bibfnamefont {W.~P.}\ \bibnamefont {Reinhardt}},\
  }\href {\doibase 10.1103/PhysRevA.62.063610} {\bibfield  {journal} {\bibinfo
  {journal} {Phys. Rev. A}\ }\textbf {\bibinfo {volume} {62}},\ \bibinfo
  {pages} {063610} (\bibinfo {year} {2000})}\BibitemShut {NoStop}%
\bibitem [{\citenamefont {Cominotti}\ \emph {et~al.}(2014)\citenamefont
  {Cominotti}, \citenamefont {Rossini}, \citenamefont {Rizzi}, \citenamefont
  {Hekking},\ and\ \citenamefont {Minguzzi}}]{Cominotti2014}%
  \BibitemOpen
  \bibfield  {author} {\bibinfo {author} {\bibfnamefont {M.}~\bibnamefont
  {Cominotti}}, \bibinfo {author} {\bibfnamefont {D.}~\bibnamefont {Rossini}},
  \bibinfo {author} {\bibfnamefont {M.}~\bibnamefont {Rizzi}}, \bibinfo
  {author} {\bibfnamefont {F.}~\bibnamefont {Hekking}}, \ and\ \bibinfo
  {author} {\bibfnamefont {A.}~\bibnamefont {Minguzzi}},\ }\href {\doibase
  10.1103/PhysRevLett.113.025301} {\bibfield  {journal} {\bibinfo  {journal}
  {Phys. Rev. Lett.}\ }\textbf {\bibinfo {volume} {113}},\ \bibinfo {pages}
  {025301} (\bibinfo {year} {2014})}\BibitemShut {NoStop}%
\bibitem [{\citenamefont {Shamriz}\ and\ \citenamefont
  {Malomed}(2018)}]{Shamriz_2018}%
  \BibitemOpen
  \bibfield  {author} {\bibinfo {author} {\bibfnamefont {E.}~\bibnamefont
  {Shamriz}}\ and\ \bibinfo {author} {\bibfnamefont {B.~A.}\ \bibnamefont
  {Malomed}},\ }\href {\doibase 10.1103/PhysRevE.98.052203} {\bibfield
  {journal} {\bibinfo  {journal} {Phys. Rev. E}\ }\textbf {\bibinfo {volume}
  {98}},\ \bibinfo {pages} {052203} (\bibinfo {year} {2018})}\BibitemShut
  {NoStop}%
\bibitem [{\citenamefont {Hakim}(1997)}]{Hakim1997}%
  \BibitemOpen
  \bibfield  {author} {\bibinfo {author} {\bibfnamefont {V.}~\bibnamefont
  {Hakim}},\ }\href {\doibase 10.1103/PhysRevE.55.2835} {\bibfield  {journal}
  {\bibinfo  {journal} {Phys. Rev. E}\ }\textbf {\bibinfo {volume} {55}},\
  \bibinfo {pages} {2835} (\bibinfo {year} {1997})}\BibitemShut {NoStop}%
\bibitem [{\citenamefont {Freire}\ \emph {et~al.}(1997)\citenamefont {Freire},
  \citenamefont {Arovas},\ and\ \citenamefont {Levine}}]{Freire1997}%
  \BibitemOpen
  \bibfield  {author} {\bibinfo {author} {\bibfnamefont {J.~A.}\ \bibnamefont
  {Freire}}, \bibinfo {author} {\bibfnamefont {D.~P.}\ \bibnamefont {Arovas}},
  \ and\ \bibinfo {author} {\bibfnamefont {H.}~\bibnamefont {Levine}},\ }\href
  {\doibase 10.1103/PhysRevLett.79.5054} {\bibfield  {journal} {\bibinfo
  {journal} {Phys. Rev. Lett.}\ }\textbf {\bibinfo {volume} {79}},\ \bibinfo
  {pages} {5054} (\bibinfo {year} {1997})}\BibitemShut {NoStop}%
\bibitem [{\citenamefont {Katsimiga}\ \emph {et~al.}(2018)\citenamefont
  {Katsimiga}, \citenamefont {Mistakidis}, \citenamefont {Koutentakis},
  \citenamefont {Kevrekidis},\ and\ \citenamefont
  {Schmelcher}}]{Katsimiga_2018}%
  \BibitemOpen
  \bibfield  {author} {\bibinfo {author} {\bibfnamefont {G.~C.}\ \bibnamefont
  {Katsimiga}}, \bibinfo {author} {\bibfnamefont {S.~I.}\ \bibnamefont
  {Mistakidis}}, \bibinfo {author} {\bibfnamefont {G.~M.}\ \bibnamefont
  {Koutentakis}}, \bibinfo {author} {\bibfnamefont {P.~G.}\ \bibnamefont
  {Kevrekidis}}, \ and\ \bibinfo {author} {\bibfnamefont {P.}~\bibnamefont
  {Schmelcher}},\ }\href {\doibase 10.1103/PhysRevA.98.013632} {\bibfield
  {journal} {\bibinfo  {journal} {Phys. Rev. A}\ }\textbf {\bibinfo {volume}
  {98}},\ \bibinfo {pages} {013632} (\bibinfo {year} {2018})}\BibitemShut
  {NoStop}%
\bibitem [{\citenamefont {Khlebnikov}(2005)}]{Khlebnikov2005a}%
  \BibitemOpen
  \bibfield  {author} {\bibinfo {author} {\bibfnamefont {S.}~\bibnamefont
  {Khlebnikov}},\ }\href {\doibase 10.1103/PhysRevA.71.013602} {\bibfield
  {journal} {\bibinfo  {journal} {Phys. Rev. A}\ }\textbf {\bibinfo {volume}
  {71}},\ \bibinfo {pages} {013602} (\bibinfo {year} {2005})}\BibitemShut
  {NoStop}%
\bibitem [{\citenamefont {Polo}\ \emph {et~al.}(2018)\citenamefont {Polo},
  \citenamefont {Ahufinger}, \citenamefont {Hekking},\ and\ \citenamefont
  {Minguzzi}}]{Polo_2018}%
  \BibitemOpen
  \bibfield  {author} {\bibinfo {author} {\bibfnamefont {J.}~\bibnamefont
  {Polo}}, \bibinfo {author} {\bibfnamefont {V.}~\bibnamefont {Ahufinger}},
  \bibinfo {author} {\bibfnamefont {F.~W.~J.}\ \bibnamefont {Hekking}}, \ and\
  \bibinfo {author} {\bibfnamefont {A.}~\bibnamefont {Minguzzi}},\ }\href
  {\doibase 10.1103/PhysRevLett.121.090404} {\bibfield  {journal} {\bibinfo
  {journal} {Phys. Rev. Lett.}\ }\textbf {\bibinfo {volume} {121}},\ \bibinfo
  {pages} {090404} (\bibinfo {year} {2018})}\BibitemShut {NoStop}%
\bibitem [{\citenamefont {Kheruntsyan}\ \emph {et~al.}(2003)\citenamefont
  {Kheruntsyan}, \citenamefont {Gangardt}, \citenamefont {Drummond},\ and\
  \citenamefont {Shlyapnikov}}]{Kheruntsyan2003}%
  \BibitemOpen
  \bibfield  {author} {\bibinfo {author} {\bibfnamefont {K.~V.}\ \bibnamefont
  {Kheruntsyan}}, \bibinfo {author} {\bibfnamefont {D.~M.}\ \bibnamefont
  {Gangardt}}, \bibinfo {author} {\bibfnamefont {P.~D.}\ \bibnamefont
  {Drummond}}, \ and\ \bibinfo {author} {\bibfnamefont {G.~V.}\ \bibnamefont
  {Shlyapnikov}},\ }\href {\doibase 10.1103/PhysRevLett.91.040403} {\bibfield
  {journal} {\bibinfo  {journal} {Phys. Rev. Lett.}\ }\textbf {\bibinfo
  {volume} {91}},\ \bibinfo {pages} {040403} (\bibinfo {year}
  {2003})}\BibitemShut {NoStop}%
\bibitem [{\citenamefont {Kumar}\ \emph {et~al.}(2018)\citenamefont {Kumar},
  \citenamefont {Dubessy}, \citenamefont {Badr}, \citenamefont {{De Rossi}},
  \citenamefont {{de Go{\"{e}}r de Herve}}, \citenamefont {Longchambon},\ and\
  \citenamefont {Perrin}}]{Kumar2018}%
  \BibitemOpen
  \bibfield  {author} {\bibinfo {author} {\bibfnamefont {A.}~\bibnamefont
  {Kumar}}, \bibinfo {author} {\bibfnamefont {R.}~\bibnamefont {Dubessy}},
  \bibinfo {author} {\bibfnamefont {T.}~\bibnamefont {Badr}}, \bibinfo {author}
  {\bibfnamefont {C.}~\bibnamefont {{De Rossi}}}, \bibinfo {author}
  {\bibfnamefont {M.}~\bibnamefont {{de Go{\"{e}}r de Herve}}}, \bibinfo
  {author} {\bibfnamefont {L.}~\bibnamefont {Longchambon}}, \ and\ \bibinfo
  {author} {\bibfnamefont {H.}~\bibnamefont {Perrin}},\ }\href {\doibase
  10.1103/PhysRevA.97.043615} {\bibfield  {journal} {\bibinfo  {journal} {Phys.
  Rev. A}\ }\textbf {\bibinfo {volume} {97}},\ \bibinfo {pages} {043615}
  (\bibinfo {year} {2018})}\BibitemShut {NoStop}%
\bibitem [{\citenamefont {Smerzi}\ \emph {et~al.}(1997)\citenamefont {Smerzi},
  \citenamefont {Fantoni}, \citenamefont {Giovanazzi},\ and\ \citenamefont
  {Shenoy}}]{Smerzi1997}%
  \BibitemOpen
  \bibfield  {author} {\bibinfo {author} {\bibfnamefont {A.}~\bibnamefont
  {Smerzi}}, \bibinfo {author} {\bibfnamefont {S.}~\bibnamefont {Fantoni}},
  \bibinfo {author} {\bibfnamefont {S.}~\bibnamefont {Giovanazzi}}, \ and\
  \bibinfo {author} {\bibfnamefont {S.~R.}\ \bibnamefont {Shenoy}},\ }\href
  {\doibase 10.1103/PhysRevLett.79.4950} {\bibfield  {journal} {\bibinfo
  {journal} {Phys. Rev. Lett.}\ }\textbf {\bibinfo {volume} {79}},\ \bibinfo
  {pages} {4950} (\bibinfo {year} {1997})}\BibitemShut {NoStop}%
\bibitem [{\citenamefont {Davis}\ \emph {et~al.}(2001)\citenamefont {Davis},
  \citenamefont {Morgan},\ and\ \citenamefont {Burnett}}]{Davis2001}%
  \BibitemOpen
  \bibfield  {author} {\bibinfo {author} {\bibfnamefont {M.~J.}\ \bibnamefont
  {Davis}}, \bibinfo {author} {\bibfnamefont {S.~A.}\ \bibnamefont {Morgan}}, \
  and\ \bibinfo {author} {\bibfnamefont {K.}~\bibnamefont {Burnett}},\ }\href
  {\doibase 10.1103/PhysRevLett.87.160402} {\bibfield  {journal} {\bibinfo
  {journal} {Phys. Rev. Lett.}\ }\textbf {\bibinfo {volume} {87}},\ \bibinfo
  {pages} {160402} (\bibinfo {year} {2001})}\BibitemShut {NoStop}%
\bibitem [{\citenamefont {Blakie}\ \emph {et~al.}(2008)\citenamefont {Blakie},
  \citenamefont {Bradley}, \citenamefont {Davis}, \citenamefont {Ballagh},\
  and\ \citenamefont {Gardiner}}]{Blakie2008}%
  \BibitemOpen
  \bibfield  {author} {\bibinfo {author} {\bibfnamefont {P.~B.}\ \bibnamefont
  {Blakie}}, \bibinfo {author} {\bibfnamefont {A.~S.}\ \bibnamefont {Bradley}},
  \bibinfo {author} {\bibfnamefont {M.~J.}\ \bibnamefont {Davis}}, \bibinfo
  {author} {\bibfnamefont {R.}~\bibnamefont {Ballagh}}, \ and\ \bibinfo
  {author} {\bibfnamefont {C.~W.}\ \bibnamefont {Gardiner}},\ }\href {\doibase
  10.1080/00018730802564254} {\bibfield  {journal} {\bibinfo  {journal}
  {Advances in Physics}\ }\textbf {\bibinfo {volume} {57}},\ \bibinfo {pages}
  {363} (\bibinfo {year} {2008})},\ \Eprint
  {http://arxiv.org/abs/https://doi.org/10.1080/00018730802564254}
  {https://doi.org/10.1080/00018730802564254} \BibitemShut {NoStop}%
\bibitem [{\citenamefont {Berloff}\ \emph {et~al.}(2014)\citenamefont
  {Berloff}, \citenamefont {Brachet},\ and\ \citenamefont
  {Proukakis}}]{Berloff2014a}%
  \BibitemOpen
  \bibfield  {author} {\bibinfo {author} {\bibfnamefont {N.~G.}\ \bibnamefont
  {Berloff}}, \bibinfo {author} {\bibfnamefont {M.}~\bibnamefont {Brachet}}, \
  and\ \bibinfo {author} {\bibfnamefont {N.~P.}\ \bibnamefont {Proukakis}},\
  }\href {\doibase 10.1073/pnas.1312549111} {\bibfield  {journal} {\bibinfo
  {journal} {Proceedings of the National Academy of Sciences}\ }\textbf
  {\bibinfo {volume} {111}},\ \bibinfo {pages} {4675} (\bibinfo {year}
  {2014})}\BibitemShut {NoStop}%
\bibitem [{\citenamefont {Girardeau}(1960)}]{Girardeau_1960}%
  \BibitemOpen
  \bibfield  {author} {\bibinfo {author} {\bibfnamefont {M.}~\bibnamefont
  {Girardeau}},\ }\href {\doibase 10.1063/1.1703687} {\bibfield  {journal}
  {\bibinfo  {journal} {J. Math. Phys.}\ }\textbf {\bibinfo {volume} {1}},\
  \bibinfo {pages} {516} (\bibinfo {year} {1960})}\BibitemShut {NoStop}%
\bibitem [{\citenamefont {Girardeau}\ and\ \citenamefont
  {Wright}(2000)}]{Wright_2000}%
  \BibitemOpen
  \bibfield  {author} {\bibinfo {author} {\bibfnamefont {M.~D.}\ \bibnamefont
  {Girardeau}}\ and\ \bibinfo {author} {\bibfnamefont {E.~M.}\ \bibnamefont
  {Wright}},\ }\href {\doibase 10.1103/PhysRevLett.84.5691} {\bibfield
  {journal} {\bibinfo  {journal} {Phys. Rev. Lett.}\ }\textbf {\bibinfo
  {volume} {84}},\ \bibinfo {pages} {5691} (\bibinfo {year}
  {2000})}\BibitemShut {NoStop}%
\bibitem [{\citenamefont {Yukalov}\ and\ \citenamefont
  {Girardeau}(2005)}]{Girardeau_2005}%
  \BibitemOpen
  \bibfield  {author} {\bibinfo {author} {\bibfnamefont {V.~I.}\ \bibnamefont
  {Yukalov}}\ and\ \bibinfo {author} {\bibfnamefont {M.~D.}\ \bibnamefont
  {Girardeau}},\ }\href {http://stacks.iop.org/1612-202X/2/i=8/a=001}
  {\bibfield  {journal} {\bibinfo  {journal} {Laser Physics Letters}\ }\textbf
  {\bibinfo {volume} {2}},\ \bibinfo {pages} {375} (\bibinfo {year}
  {2005})}\BibitemShut {NoStop}%
\bibitem [{Note1()}]{Note1}%
  \BibitemOpen
  \bibinfo {note} {The calculation for $\ell =2$ yields no qualitative
  difference in the current dynamics, just a faster decay of the
  oscillations.}\BibitemShut {Stop}%
\bibitem [{\citenamefont {Albiez}\ \emph {et~al.}(2005)\citenamefont {Albiez},
  \citenamefont {Gati}, \citenamefont {F{\"{o}}lling}, \citenamefont
  {Hunsmann}, \citenamefont {Cristiani},\ and\ \citenamefont
  {Oberthaler}}]{Albiez2005}%
  \BibitemOpen
  \bibfield  {author} {\bibinfo {author} {\bibfnamefont {M.}~\bibnamefont
  {Albiez}}, \bibinfo {author} {\bibfnamefont {R.}~\bibnamefont {Gati}},
  \bibinfo {author} {\bibfnamefont {J.}~\bibnamefont {F{\"{o}}lling}}, \bibinfo
  {author} {\bibfnamefont {S.}~\bibnamefont {Hunsmann}}, \bibinfo {author}
  {\bibfnamefont {M.}~\bibnamefont {Cristiani}}, \ and\ \bibinfo {author}
  {\bibfnamefont {M.~K.}\ \bibnamefont {Oberthaler}},\ }\href {\doibase
  10.1103/PhysRevLett.95.010402} {\bibfield  {journal} {\bibinfo  {journal}
  {Phys. Rev. Lett.}\ }\textbf {\bibinfo {volume} {95}},\ \bibinfo {pages}
  {010402} (\bibinfo {year} {2005})}\BibitemShut {NoStop}%
\bibitem [{sup()}]{supplemental}%
  \BibitemOpen
  \href@noop {} {}\bibinfo {note} {See Supplemental Material for details, which
  includes
  Refs.~\cite{Brachet2012,Gardiner2002a,Gardiner2003a,Krstulovic2011a,Millard_1969,Wright_2002,Ananikian_2006,Nesterenko_2014}}\BibitemShut
  {NoStop}%
\bibitem [{\citenamefont {Caradoc-Davies}\ \emph {et~al.}(1999)\citenamefont
  {Caradoc-Davies}, \citenamefont {Ballagh},\ and\ \citenamefont
  {Burnett}}]{Caradoc-Davies1999}%
  \BibitemOpen
  \bibfield  {author} {\bibinfo {author} {\bibfnamefont {B.~M.}\ \bibnamefont
  {Caradoc-Davies}}, \bibinfo {author} {\bibfnamefont {R.~J.}\ \bibnamefont
  {Ballagh}}, \ and\ \bibinfo {author} {\bibfnamefont {K.}~\bibnamefont
  {Burnett}},\ }\href {\doibase 10.1103/PhysRevLett.83.895} {\bibfield
  {journal} {\bibinfo  {journal} {Phys. Rev. Lett.}\ }\textbf {\bibinfo
  {volume} {83}},\ \bibinfo {pages} {895} (\bibinfo {year} {1999})}\BibitemShut
  {NoStop}%
\bibitem [{Den()}]{DensityDev}%
  \BibitemOpen
  \href@noop {} {}\bibinfo {note} {For the sake of clarity we plot only the
  density deviations with respect to the time-averaged value.}\BibitemShut
  {Stop}%
\bibitem [{Fit()}]{Fitting}%
  \BibitemOpen
  \href@noop {} {}\bibinfo {note} {We fit all the data sets with the same
  model: $J(t)=Ae^{-\Gamma_At}+B\cos{[\omega t+\phi]}e^{-\Gamma_Bt}$ that
  allows to capture both a simple exponential decay and damped oscillations.
  This model works also remarkably well in the intermediate regime where the
  two effects are simultaneously present.}\BibitemShut {Stop}%
\bibitem [{\citenamefont {Karpiuk}\ \emph {et~al.}(2012)\citenamefont
  {Karpiuk}, \citenamefont {Deuar}, \citenamefont {Bienias}, \citenamefont
  {Witkowska}, \citenamefont {Paw{\l}owski}, \citenamefont {Gajda},
  \citenamefont {Rz{\c{a}}{\.{z}}ewski},\ and\ \citenamefont
  {Brewczyk}}]{Karpiuk2012}%
  \BibitemOpen
  \bibfield  {author} {\bibinfo {author} {\bibfnamefont {T.}~\bibnamefont
  {Karpiuk}}, \bibinfo {author} {\bibfnamefont {P.}~\bibnamefont {Deuar}},
  \bibinfo {author} {\bibfnamefont {P.}~\bibnamefont {Bienias}}, \bibinfo
  {author} {\bibfnamefont {E.}~\bibnamefont {Witkowska}}, \bibinfo {author}
  {\bibfnamefont {K.}~\bibnamefont {Paw{\l}owski}}, \bibinfo {author}
  {\bibfnamefont {M.}~\bibnamefont {Gajda}}, \bibinfo {author} {\bibfnamefont
  {K.}~\bibnamefont {Rz{\c{a}}{\.{z}}ewski}}, \ and\ \bibinfo {author}
  {\bibfnamefont {M.}~\bibnamefont {Brewczyk}},\ }\href {\doibase
  10.1103/PhysRevLett.109.205302} {\bibfield  {journal} {\bibinfo  {journal}
  {Phys. Rev. Lett.}\ }\textbf {\bibinfo {volume} {109}},\ \bibinfo {pages}
  {205302} (\bibinfo {year} {2012})}\BibitemShut {NoStop}%
\bibitem [{\citenamefont {Bilas}\ and\ \citenamefont
  {Pavloff}(2005)}]{Bilas2005}%
  \BibitemOpen
  \bibfield  {author} {\bibinfo {author} {\bibfnamefont {N.}~\bibnamefont
  {Bilas}}\ and\ \bibinfo {author} {\bibfnamefont {N.}~\bibnamefont
  {Pavloff}},\ }\href {\doibase 10.1103/PhysRevA.72.033618} {\bibfield
  {journal} {\bibinfo  {journal} {Phys. Rev. A}\ }\textbf {\bibinfo {volume}
  {72}},\ \bibinfo {pages} {033618} (\bibinfo {year} {2005})}\BibitemShut
  {NoStop}%
\bibitem [{\citenamefont {{Astafiev}}\ \emph {et~al.}(2012)\citenamefont
  {{Astafiev}}, \citenamefont {{Ioffe}}, \citenamefont {{Kafanov}},
  \citenamefont {{Pashkin}}, \citenamefont {{Arutyunov}}, \citenamefont
  {{Shahar}}, \citenamefont {{Cohen}},\ and\ \citenamefont
  {{Tsai}}}]{Astafiev_2012}%
  \BibitemOpen
  \bibfield  {author} {\bibinfo {author} {\bibfnamefont {O.~V.}\ \bibnamefont
  {{Astafiev}}}, \bibinfo {author} {\bibfnamefont {L.~B.}\ \bibnamefont
  {{Ioffe}}}, \bibinfo {author} {\bibfnamefont {S.}~\bibnamefont {{Kafanov}}},
  \bibinfo {author} {\bibfnamefont {Y.~A.}\ \bibnamefont {{Pashkin}}}, \bibinfo
  {author} {\bibfnamefont {K.~Y.}\ \bibnamefont {{Arutyunov}}}, \bibinfo
  {author} {\bibfnamefont {D.}~\bibnamefont {{Shahar}}}, \bibinfo {author}
  {\bibfnamefont {O.}~\bibnamefont {{Cohen}}}, \ and\ \bibinfo {author}
  {\bibfnamefont {J.~S.}\ \bibnamefont {{Tsai}}},\ }\href {\doibase
  10.1038/nature10930} {\bibfield  {journal} {\bibinfo  {journal} {Nature}\
  }\textbf {\bibinfo {volume} {484}},\ \bibinfo {pages} {355} (\bibinfo {year}
  {2012})}\BibitemShut {NoStop}%
\bibitem [{\citenamefont {Mooij}\ and\ \citenamefont
  {Nazarov}(2006)}]{Moij2006}%
  \BibitemOpen
  \bibfield  {author} {\bibinfo {author} {\bibfnamefont {J.~E.}\ \bibnamefont
  {Mooij}}\ and\ \bibinfo {author} {\bibfnamefont {Y.~V.}\ \bibnamefont
  {Nazarov}},\ }\href@noop {} {\bibfield  {journal} {\bibinfo  {journal}
  {Nature Phys.}\ }\textbf {\bibinfo {volume} {2}},\ \bibinfo {pages} {169}
  (\bibinfo {year} {2006})}\BibitemShut {NoStop}%
\bibitem [{\citenamefont {Corman}\ \emph {et~al.}(2014)\citenamefont {Corman},
  \citenamefont {Chomaz}, \citenamefont {Bienaim\'e}, \citenamefont
  {Desbuquois}, \citenamefont {Weitenberg}, \citenamefont {Nascimb\`ene},
  \citenamefont {Dalibard},\ and\ \citenamefont {Beugnon}}]{Corman2014}%
  \BibitemOpen
  \bibfield  {author} {\bibinfo {author} {\bibfnamefont {L.}~\bibnamefont
  {Corman}}, \bibinfo {author} {\bibfnamefont {L.}~\bibnamefont {Chomaz}},
  \bibinfo {author} {\bibfnamefont {T.}~\bibnamefont {Bienaim\'e}}, \bibinfo
  {author} {\bibfnamefont {R.}~\bibnamefont {Desbuquois}}, \bibinfo {author}
  {\bibfnamefont {C.}~\bibnamefont {Weitenberg}}, \bibinfo {author}
  {\bibfnamefont {S.}~\bibnamefont {Nascimb\`ene}}, \bibinfo {author}
  {\bibfnamefont {J.}~\bibnamefont {Dalibard}}, \ and\ \bibinfo {author}
  {\bibfnamefont {J.}~\bibnamefont {Beugnon}},\ }\href {\doibase
  10.1103/PhysRevLett.113.135302} {\bibfield  {journal} {\bibinfo  {journal}
  {Phys. Rev. Lett.}\ }\textbf {\bibinfo {volume} {113}},\ \bibinfo {pages}
  {135302} (\bibinfo {year} {2014})}\BibitemShut {NoStop}%
\bibitem [{\citenamefont {Eckel}\ \emph
  {et~al.}(2014{\natexlab{b}})\citenamefont {Eckel}, \citenamefont
  {Jendrzejewski}, \citenamefont {Kumar}, \citenamefont {Lobb},\ and\
  \citenamefont {Campbell}}]{Eckel2014a}%
  \BibitemOpen
  \bibfield  {author} {\bibinfo {author} {\bibfnamefont {S.}~\bibnamefont
  {Eckel}}, \bibinfo {author} {\bibfnamefont {F.}~\bibnamefont
  {Jendrzejewski}}, \bibinfo {author} {\bibfnamefont {A.}~\bibnamefont
  {Kumar}}, \bibinfo {author} {\bibfnamefont {C.~J.}\ \bibnamefont {Lobb}}, \
  and\ \bibinfo {author} {\bibfnamefont {G.~K.}\ \bibnamefont {Campbell}},\
  }\href {\doibase 10.1103/PhysRevX.4.031052} {\bibfield  {journal} {\bibinfo
  {journal} {Phys. Rev. X}\ }\textbf {\bibinfo {volume} {4}},\ \bibinfo {pages}
  {031052} (\bibinfo {year} {2014}{\natexlab{b}})}\BibitemShut {NoStop}%
\bibitem [{\citenamefont {Marti}\ \emph {et~al.}(2015)\citenamefont {Marti},
  \citenamefont {Olf},\ and\ \citenamefont {Stamper-Kurn}}]{Marti2014}%
  \BibitemOpen
  \bibfield  {author} {\bibinfo {author} {\bibfnamefont {G.~E.}\ \bibnamefont
  {Marti}}, \bibinfo {author} {\bibfnamefont {R.}~\bibnamefont {Olf}}, \ and\
  \bibinfo {author} {\bibfnamefont {D.~M.}\ \bibnamefont {Stamper-Kurn}},\
  }\href {\doibase 10.1103/PhysRevA.91.013602} {\bibfield  {journal} {\bibinfo
  {journal} {Phys. Rev. A}\ }\textbf {\bibinfo {volume} {91}},\ \bibinfo
  {pages} {013602} (\bibinfo {year} {2015})}\BibitemShut {NoStop}%
\bibitem [{\citenamefont {Kumar}\ \emph {et~al.}(2016)\citenamefont {Kumar},
  \citenamefont {Anderson}, \citenamefont {Phillips}, \citenamefont {Eckel},
  \citenamefont {Campbell},\ and\ \citenamefont {Stringari}}]{Kumar2016}%
  \BibitemOpen
  \bibfield  {author} {\bibinfo {author} {\bibfnamefont {A.}~\bibnamefont
  {Kumar}}, \bibinfo {author} {\bibfnamefont {N.}~\bibnamefont {Anderson}},
  \bibinfo {author} {\bibfnamefont {W.~D.}\ \bibnamefont {Phillips}}, \bibinfo
  {author} {\bibfnamefont {S.}~\bibnamefont {Eckel}}, \bibinfo {author}
  {\bibfnamefont {G.~K.}\ \bibnamefont {Campbell}}, \ and\ \bibinfo {author}
  {\bibfnamefont {S.}~\bibnamefont {Stringari}},\ }\href {\doibase
  10.1088/1367-2630/18/2/025001} {\bibfield  {journal} {\bibinfo  {journal}
  {New Journal of Physics}\ }\textbf {\bibinfo {volume} {18}},\ \bibinfo
  {pages} {25001} (\bibinfo {year} {2016})}\BibitemShut {NoStop}%
\bibitem [{\citenamefont {Brachet}(2012)}]{Brachet2012}%
  \BibitemOpen
  \bibfield  {author} {\bibinfo {author} {\bibfnamefont {M.}~\bibnamefont
  {Brachet}},\ }\href {\doibase 10.1016/j.crhy.2012.10.006} {\bibfield
  {journal} {\bibinfo  {journal} {Comptes Rendus Physique}\ }\textbf {\bibinfo
  {volume} {13}},\ \bibinfo {pages} {954} (\bibinfo {year} {2012})}\BibitemShut
  {NoStop}%
\bibitem [{\citenamefont {Gardiner}\ \emph {et~al.}(2002)\citenamefont
  {Gardiner}, \citenamefont {Anglin},\ and\ \citenamefont
  {Fudge}}]{Gardiner2002a}%
  \BibitemOpen
  \bibfield  {author} {\bibinfo {author} {\bibfnamefont {C.~W.}\ \bibnamefont
  {Gardiner}}, \bibinfo {author} {\bibfnamefont {J.~R.}\ \bibnamefont
  {Anglin}}, \ and\ \bibinfo {author} {\bibfnamefont {T.~I.~A.}\ \bibnamefont
  {Fudge}},\ }\href {\doibase 10.1088/0953-4075/35/6/310} {\bibfield  {journal}
  {\bibinfo  {journal} {J. Phys. B}\ }\textbf {\bibinfo {volume} {35}},\
  \bibinfo {pages} {1555} (\bibinfo {year} {2002})}\BibitemShut {NoStop}%
\bibitem [{\citenamefont {Gardiner}\ and\ \citenamefont
  {Davis}(2003)}]{Gardiner2003a}%
  \BibitemOpen
  \bibfield  {author} {\bibinfo {author} {\bibfnamefont {C.~W.}\ \bibnamefont
  {Gardiner}}\ and\ \bibinfo {author} {\bibfnamefont {M.~J.}\ \bibnamefont
  {Davis}},\ }\href {\doibase 10.1088/0953-4075/36/23/010} {\bibfield
  {journal} {\bibinfo  {journal} {J. Phys. B}\ }\textbf {\bibinfo {volume}
  {36}},\ \bibinfo {pages} {4731} (\bibinfo {year} {2003})}\BibitemShut
  {NoStop}%
\bibitem [{\citenamefont {Krstulovic}\ and\ \citenamefont
  {Brachet}(2011)}]{Krstulovic2011a}%
  \BibitemOpen
  \bibfield  {author} {\bibinfo {author} {\bibfnamefont {G.}~\bibnamefont
  {Krstulovic}}\ and\ \bibinfo {author} {\bibfnamefont {M.}~\bibnamefont
  {Brachet}},\ }\href {\doibase 10.1103/PhysRevE.83.066311} {\bibfield
  {journal} {\bibinfo  {journal} {Phys. Rev. E}\ }\textbf {\bibinfo {volume}
  {83}},\ \bibinfo {pages} {066311} (\bibinfo {year} {2011})}\BibitemShut
  {NoStop}%
\bibitem [{\citenamefont {Millard}(1969)}]{Millard_1969}%
  \BibitemOpen
  \bibfield  {author} {\bibinfo {author} {\bibfnamefont {K.}~\bibnamefont
  {Millard}},\ }\href {\doibase 10.1063/1.1664762} {\bibfield  {journal}
  {\bibinfo  {journal} {J. Math. Phys.}\ }\textbf {\bibinfo {volume} {10}},\
  \bibinfo {pages} {7} (\bibinfo {year} {1969})}\BibitemShut {NoStop}%
\bibitem [{\citenamefont {Das}\ \emph {et~al.}(2002)\citenamefont {Das},
  \citenamefont {Girardeau},\ and\ \citenamefont {Wright}}]{Wright_2002}%
  \BibitemOpen
  \bibfield  {author} {\bibinfo {author} {\bibfnamefont {K.~K.}\ \bibnamefont
  {Das}}, \bibinfo {author} {\bibfnamefont {M.~D.}\ \bibnamefont {Girardeau}},
  \ and\ \bibinfo {author} {\bibfnamefont {E.~M.}\ \bibnamefont {Wright}},\
  }\href {\doibase 10.1103/PhysRevLett.89.170404} {\bibfield  {journal}
  {\bibinfo  {journal} {Phys. Rev. Lett.}\ }\textbf {\bibinfo {volume} {89}},\
  \bibinfo {pages} {170404} (\bibinfo {year} {2002})}\BibitemShut {NoStop}%
\bibitem [{\citenamefont {Ananikian}\ and\ \citenamefont
  {Bergeman}(2006)}]{Ananikian_2006}%
  \BibitemOpen
  \bibfield  {author} {\bibinfo {author} {\bibfnamefont {D.}~\bibnamefont
  {Ananikian}}\ and\ \bibinfo {author} {\bibfnamefont {T.}~\bibnamefont
  {Bergeman}},\ }\href {\doibase 10.1103/PhysRevA.73.013604} {\bibfield
  {journal} {\bibinfo  {journal} {Phys. Rev. A}\ }\textbf {\bibinfo {volume}
  {73}},\ \bibinfo {pages} {013604} (\bibinfo {year} {2006})}\BibitemShut
  {NoStop}%
\bibitem [{\citenamefont {Nesterenko}\ \emph {et~al.}(2014)\citenamefont
  {Nesterenko}, \citenamefont {Novikov},\ and\ \citenamefont
  {Suraud}}]{Nesterenko_2014}%
  \BibitemOpen
  \bibfield  {author} {\bibinfo {author} {\bibfnamefont {V.~O.}\ \bibnamefont
  {Nesterenko}}, \bibinfo {author} {\bibfnamefont {A.~N.}\ \bibnamefont
  {Novikov}}, \ and\ \bibinfo {author} {\bibfnamefont {E.}~\bibnamefont
  {Suraud}},\ }\href {\doibase 10.1088/1054-660x/24/12/125501} {\bibfield
  {journal} {\bibinfo  {journal} {Laser Physics}\ }\textbf {\bibinfo {volume}
  {24}},\ \bibinfo {pages} {125501} (\bibinfo {year} {2014})}\BibitemShut
  {NoStop}%
\end{thebibliography}%


\begin{thebibliography}{14}%
\makeatletter
\providecommand \@ifxundefined [1]{%
 \@ifx{#1\undefined}
}%
\providecommand \@ifnum [1]{%
 \ifnum #1\expandafter \@firstoftwo
 \else \expandafter \@secondoftwo
 \fi
}%
\providecommand \@ifx [1]{%
 \ifx #1\expandafter \@firstoftwo
 \else \expandafter \@secondoftwo
 \fi
}%
\providecommand \natexlab [1]{#1}%
\providecommand \enquote  [1]{``#1''}%
\providecommand \bibnamefont  [1]{#1}%
\providecommand \bibfnamefont [1]{#1}%
\providecommand \citenamefont [1]{#1}%
\providecommand \href@noop [0]{\@secondoftwo}%
\providecommand \href [0]{\begingroup \@sanitize@url \@href}%
\providecommand \@href[1]{\@@startlink{#1}\@@href}%
\providecommand \@@href[1]{\endgroup#1\@@endlink}%
\providecommand \@sanitize@url [0]{\catcode `\\12\catcode `\$12\catcode
  `\&12\catcode `\#12\catcode `\^12\catcode `\_12\catcode `\%12\relax}%
\providecommand \@@startlink[1]{}%
\providecommand \@@endlink[0]{}%
\providecommand \url  [0]{\begingroup\@sanitize@url \@url }%
\providecommand \@url [1]{\endgroup\@href {#1}{\urlprefix }}%
\providecommand \urlprefix  [0]{URL }%
\providecommand \Eprint [0]{\href }%
\providecommand \doibase [0]{http://dx.doi.org/}%
\providecommand \selectlanguage [0]{\@gobble}%
\providecommand \bibinfo  [0]{\@secondoftwo}%
\providecommand \bibfield  [0]{\@secondoftwo}%
\providecommand \translation [1]{[#1]}%
\providecommand \BibitemOpen [0]{}%
\providecommand \bibitemStop [0]{}%
\providecommand \bibitemNoStop [0]{.\EOS\space}%
\providecommand \EOS [0]{\spacefactor3000\relax}%
\providecommand \BibitemShut  [1]{\csname bibitem#1\endcsname}%
\let\auto@bib@innerbib\@empty
\bibitem [{\citenamefont {Smerzi}\ \emph {et~al.}(1997)\citenamefont {Smerzi},
  \citenamefont {Fantoni}, \citenamefont {Giovanazzi},\ and\ \citenamefont
  {Shenoy}}]{Smerzi1997}%
  \BibitemOpen
  \bibfield  {author} {\bibinfo {author} {\bibfnamefont {A.}~\bibnamefont
  {Smerzi}}, \bibinfo {author} {\bibfnamefont {S.}~\bibnamefont {Fantoni}},
  \bibinfo {author} {\bibfnamefont {S.}~\bibnamefont {Giovanazzi}}, \ and\
  \bibinfo {author} {\bibfnamefont {S.~R.}\ \bibnamefont {Shenoy}},\ }\href
  {\doibase 10.1103/PhysRevLett.79.4950} {\bibfield  {journal} {\bibinfo
  {journal} {Phys. Rev. Lett.}\ }\textbf {\bibinfo {volume} {79}},\ \bibinfo
  {pages} {4950} (\bibinfo {year} {1997})}\BibitemShut {NoStop}%
\bibitem [{\citenamefont {Davis}\ \emph {et~al.}(2001)\citenamefont {Davis},
  \citenamefont {Morgan},\ and\ \citenamefont {Burnett}}]{Davis2001}%
  \BibitemOpen
  \bibfield  {author} {\bibinfo {author} {\bibfnamefont {M.~J.}\ \bibnamefont
  {Davis}}, \bibinfo {author} {\bibfnamefont {S.~A.}\ \bibnamefont {Morgan}}, \
  and\ \bibinfo {author} {\bibfnamefont {K.}~\bibnamefont {Burnett}},\ }\href
  {\doibase 10.1103/PhysRevLett.87.160402} {\bibfield  {journal} {\bibinfo
  {journal} {Phys. Rev. Lett.}\ }\textbf {\bibinfo {volume} {87}},\ \bibinfo
  {pages} {1} (\bibinfo {year} {2001})}\BibitemShut {NoStop}%
\bibitem [{\citenamefont {Berloff}\ \emph {et~al.}(2014)\citenamefont
  {Berloff}, \citenamefont {Brachet},\ and\ \citenamefont
  {Proukakis}}]{Berloff2014a}%
  \BibitemOpen
  \bibfield  {author} {\bibinfo {author} {\bibfnamefont {N.~G.}\ \bibnamefont
  {Berloff}}, \bibinfo {author} {\bibfnamefont {M.}~\bibnamefont {Brachet}}, \
  and\ \bibinfo {author} {\bibfnamefont {N.~P.}\ \bibnamefont {Proukakis}},\
  }\href {\doibase 10.1073/pnas.1312549111} {\bibfield  {journal} {\bibinfo
  {journal} {Proceedings of the National Academy of Sciences}\ }\textbf
  {\bibinfo {volume} {111}},\ \bibinfo {pages} {4675} (\bibinfo {year}
  {2014})}\BibitemShut {NoStop}%
\bibitem [{\citenamefont {Brachet}(2012)}]{Brachet2012}%
  \BibitemOpen
  \bibfield  {author} {\bibinfo {author} {\bibfnamefont {M.}~\bibnamefont
  {Brachet}},\ }\href {\doibase 10.1016/j.crhy.2012.10.006} {\bibfield
  {journal} {\bibinfo  {journal} {Comptes Rendus Physique}\ }\textbf {\bibinfo
  {volume} {13}},\ \bibinfo {pages} {954} (\bibinfo {year} {2012})}\BibitemShut
  {NoStop}%
\bibitem [{\citenamefont {Gardiner}\ \emph {et~al.}(2002)\citenamefont
  {Gardiner}, \citenamefont {Anglin},\ and\ \citenamefont
  {Fudge}}]{Gardiner2002a}%
  \BibitemOpen
  \bibfield  {author} {\bibinfo {author} {\bibfnamefont {C.~W.}\ \bibnamefont
  {Gardiner}}, \bibinfo {author} {\bibfnamefont {J.~R.}\ \bibnamefont
  {Anglin}}, \ and\ \bibinfo {author} {\bibfnamefont {T.~I.~A.}\ \bibnamefont
  {Fudge}},\ }\href {\doibase 10.1088/0953-4075/35/6/310} {\bibfield  {journal}
  {\bibinfo  {journal} {J. Phys. B}\ }\textbf {\bibinfo {volume} {35}},\
  \bibinfo {pages} {1555} (\bibinfo {year} {2002})}\BibitemShut {NoStop}%
\bibitem [{\citenamefont {Gardiner}\ and\ \citenamefont
  {Davis}(2003)}]{Gardiner2003a}%
  \BibitemOpen
  \bibfield  {author} {\bibinfo {author} {\bibfnamefont {C.~W.}\ \bibnamefont
  {Gardiner}}\ and\ \bibinfo {author} {\bibfnamefont {M.~J.}\ \bibnamefont
  {Davis}},\ }\href {\doibase 10.1088/0953-4075/36/23/010} {\bibfield
  {journal} {\bibinfo  {journal} {J. Phys. B}\ }\textbf {\bibinfo {volume}
  {36}},\ \bibinfo {pages} {4731} (\bibinfo {year} {2003})}\BibitemShut
  {NoStop}%
\bibitem [{\citenamefont {Krstulovic}\ and\ \citenamefont
  {Brachet}(2011)}]{Krstulovic2011a}%
  \BibitemOpen
  \bibfield  {author} {\bibinfo {author} {\bibfnamefont {G.}~\bibnamefont
  {Krstulovic}}\ and\ \bibinfo {author} {\bibfnamefont {M.}~\bibnamefont
  {Brachet}},\ }\href {\doibase 10.1103/PhysRevE.83.066311} {\bibfield
  {journal} {\bibinfo  {journal} {Phys. Rev. E}\ }\textbf {\bibinfo {volume}
  {83}},\ \bibinfo {pages} {066311} (\bibinfo {year} {2011})}\BibitemShut
  {NoStop}%
\bibitem [{\citenamefont {Girardeau}(1960)}]{Girardeau_1960}%
  \BibitemOpen
  \bibfield  {author} {\bibinfo {author} {\bibfnamefont {M.}~\bibnamefont
  {Girardeau}},\ }\href {\doibase 10.1063/1.1703687} {\bibfield  {journal}
  {\bibinfo  {journal} {J. Math. Phys.}\ }\textbf {\bibinfo {volume} {1}},\
  \bibinfo {pages} {516} (\bibinfo {year} {1960})}\BibitemShut {NoStop}%
\bibitem [{\citenamefont {Girardeau}\ and\ \citenamefont
  {Wright}(2000)}]{Wright_2000}%
  \BibitemOpen
  \bibfield  {author} {\bibinfo {author} {\bibfnamefont {M.~D.}\ \bibnamefont
  {Girardeau}}\ and\ \bibinfo {author} {\bibfnamefont {E.~M.}\ \bibnamefont
  {Wright}},\ }\href {\doibase 10.1103/PhysRevLett.84.5691} {\bibfield
  {journal} {\bibinfo  {journal} {Phys. Rev. Lett.}\ }\textbf {\bibinfo
  {volume} {84}},\ \bibinfo {pages} {5691} (\bibinfo {year}
  {2000})}\BibitemShut {NoStop}%
\bibitem [{\citenamefont {Yukalov}\ and\ \citenamefont
  {Girardeau}(2005)}]{Girardeau_2005}%
  \BibitemOpen
  \bibfield  {author} {\bibinfo {author} {\bibfnamefont {V.~I.}\ \bibnamefont
  {Yukalov}}\ and\ \bibinfo {author} {\bibfnamefont {M.~D.}\ \bibnamefont
  {Girardeau}},\ }\href {http://stacks.iop.org/1612-202X/2/i=8/a=001}
  {\bibfield  {journal} {\bibinfo  {journal} {Laser Physics Letters}\ }\textbf
  {\bibinfo {volume} {2}},\ \bibinfo {pages} {375} (\bibinfo {year}
  {2005})}\BibitemShut {NoStop}%
\bibitem [{\citenamefont {Millard}(1969)}]{Millard_1969}%
  \BibitemOpen
  \bibfield  {author} {\bibinfo {author} {\bibfnamefont {K.}~\bibnamefont
  {Millard}},\ }\href {\doibase 10.1063/1.1664762} {\bibfield  {journal}
  {\bibinfo  {journal} {J. Math. Phys.}\ }\textbf {\bibinfo {volume} {10}},\
  \bibinfo {pages} {7} (\bibinfo {year} {1969})}\BibitemShut {NoStop}%
\bibitem [{\citenamefont {Das}\ \emph {et~al.}(2002)\citenamefont {Das},
  \citenamefont {Girardeau},\ and\ \citenamefont {Wright}}]{Wright_2002}%
  \BibitemOpen
  \bibfield  {author} {\bibinfo {author} {\bibfnamefont {K.~K.}\ \bibnamefont
  {Das}}, \bibinfo {author} {\bibfnamefont {M.~D.}\ \bibnamefont {Girardeau}},
  \ and\ \bibinfo {author} {\bibfnamefont {E.~M.}\ \bibnamefont {Wright}},\
  }\href {\doibase 10.1103/PhysRevLett.89.170404} {\bibfield  {journal}
  {\bibinfo  {journal} {Phys. Rev. Lett.}\ }\textbf {\bibinfo {volume} {89}},\
  \bibinfo {pages} {170404} (\bibinfo {year} {2002})}\BibitemShut {NoStop}%
\bibitem [{\citenamefont {Ananikian}\ and\ \citenamefont
  {Bergeman}(2006)}]{Ananikian_2006}%
  \BibitemOpen
  \bibfield  {author} {\bibinfo {author} {\bibfnamefont {D.}~\bibnamefont
  {Ananikian}}\ and\ \bibinfo {author} {\bibfnamefont {T.}~\bibnamefont
  {Bergeman}},\ }\href {\doibase 10.1103/PhysRevA.73.013604} {\bibfield
  {journal} {\bibinfo  {journal} {Phys. Rev. A}\ }\textbf {\bibinfo {volume}
  {73}},\ \bibinfo {pages} {013604} (\bibinfo {year} {2006})}\BibitemShut
  {NoStop}%
\bibitem [{\citenamefont {Nesterenko}\ \emph {et~al.}(2014)\citenamefont
  {Nesterenko}, \citenamefont {Novikov},\ and\ \citenamefont
  {Suraud}}]{Nesterenko_2014}%
  \BibitemOpen
  \bibfield  {author} {\bibinfo {author} {\bibfnamefont {V.~O.}\ \bibnamefont
  {Nesterenko}}, \bibinfo {author} {\bibfnamefont {A.~N.}\ \bibnamefont
  {Novikov}}, \ and\ \bibinfo {author} {\bibfnamefont {E.}~\bibnamefont
  {Suraud}},\ }\href {\doibase 10.1088/1054-660x/24/12/125501} {\bibfield
  {journal} {\bibinfo  {journal} {Laser Physics}\ }\textbf {\bibinfo {volume}
  {24}},\ \bibinfo {pages} {125501} (\bibinfo {year} {2014})}\BibitemShut
  {NoStop}%
\end{thebibliography}%

\end{document}


\title{Supplemental material for:\\ Oscillations and decay of superfluid currents in a one-dimensional Bose gas on a ring}

\author{Juan Polo}
\affiliation{Univ. Grenoble Alpes, CNRS, LPMMC, 38000 Grenoble, France}
\affiliation{Quantum Systems Unit, Okinawa Institute of Science and Technology Graduate University, Onna, Okinawa 904-0495, Japan}
\author{Romain Dubessy}
\author{Paolo Pedri}
\author{H\'el\`ene Perrin}
\affiliation{Laboratoire de physique des lasers, CNRS, Universit\'e Paris 13,
Sorbonne Paris Cit\'e, 99 avenue J.-B. Cl\'ement, F-93430 Villetaneuse, France}
\author{Anna Minguzzi$^{1}$}

\date{\today}

\maketitle
\tableofcontents

\section{Methods}
We provide in this section the details of the theoretical methods used in the main text.

\paragraph{Gross-Pitaevskii equation}
We describe weakly interacting bosons, ie for $\gamma\ll 1$,  at zero temperature by the mean-field Gross-Pitaevskii equation:
\[
i\hbar\frac{\partial\psi}{\partial t}=\left(-\frac{\hbar^2}{2m}\frac{\partial^2}{\partial x^2}+V(x)+g|\psi|^2\right)\psi.
\]

\paragraph{Two-mode model}
Within the mean-field approach at zero temperature, we  model the dynamics of the system after the quench by focusing on a subspace of the occupied spatial modes $\varphi_j(x)$. In particular, for weak barrier strengths, for which the Bose gas occupies mostly  two modes, we use a two-mode approximation in which the wavefunction can be written as $\psi (x,t) = \phi_1(t)\varphi_1(x)+\phi_2(t)\varphi_2(x)$ \cite{Smerzi1997}. The amplitudes $\phi_i(t)$ of each of these two modes correspond to the occupation of the corresponding angular momentum states, while $\varphi_{1(2)}(x)$, which are built as a symmetric (antisymmetric) combination of the first and second excited states, produce states with $\pm \ell$ angular momentum. Note however that, at difference from Ref.\cite{Smerzi1997}, both modes $\varphi_j(x)$ coexist in space and their orthogonality is given by their relative phase and not by their physical spatial separation.

This simplified model allows us to describe the dynamics of the current in terms of the Josephson equations for the dual variables with respect to the usual case:  in place of relative number and phase among two macroscopic phase-coherent objects, we obtain here the Josephson oscillations of the average current  and its conjugate phase on a ring. A detailed derivation of the Josephson equations is provided in Sec.\ref{app:TMM} below.

\paragraph{Projected Gross-Pitaevskii equation}
To model the finite temperature weakly interacting regime $\gamma\ll1$ we use the classical field methodology and more precisely the projected Gross-Pitaevskii equation (PGPE) \cite{Davis2001,Berloff2014a}:
\[
i\hbar\frac{\partial\psi_C}{\partial t}=P_C\left[\left(-\frac{\hbar^2}{2m}\frac{\partial^2}{\partial x^2}+P_C\left[V(x)+g|\psi_C|^2\right]\right)\psi_C\right],
\]
where $\psi_C$ is the classical field, obtained by restricting the system to the highly populated modes only (which can be treated classically) and $P_C[...]$ is the projector onto this subspace. This projector is implemented in the single particle momentum basis, by defining a cutoff on the wavevectors $k$.
We follow the rule $k_{\rm cut}\leq2k_{\rm grid}/3$ to avoid aliasing and enforce momentum conservation (in the absence of the barrier), even in the presence of a non negligible thermal fraction \cite{Brachet2012}.
To prepare the initial state we sample an equilibrium thermal state using a stochastic PGPE \cite{Gardiner2002a,Gardiner2003a} while fixing the average number of particles \cite{Krstulovic2011a}. The simulation is run typically 100 times with different initial states and measured quantities are averaged over this ensemble. The PGPE is particularly relevant to model 1D Bose gases at finite temperature, in the quasi-condensate regime, allowing to perform quantitative comparison with experiments~\cite{Berloff2014a}.

\paragraph{Tonks-Girardeau regime}
To describe the limit of strongly interacting bosons, $\gamma\gg 1$, we focus on the Tonks-Girardeau (TG) limit of infinitely strong repulsive interactions, and  
model the dynamics by using the exact TG solution~\cite{Girardeau_1960}. In particular, we make use of the time-dependent Bose-Fermi mapping \cite{Girardeau_1960,Wright_2000,Girardeau_2005}, where the many-body wavefunction $\Psi_{TG}$ reads
\begin{equation}
\Psi_{TG}(x_1,...,x_N)=\Pi_{1\le j< \ell\le N}\text{sgn}(x_j-x_\ell)\, \text{det}[\psi_{k}(x_j,t)],\label{eq:TG_wavefunction}
\end{equation}
where $\psi_j(x,t)$ is the single-particle solution of the Schr\"odinger equation
$ i \hbar \partial_t \psi_j=\left[-\hbar^2 \partial_x^2/2m + V(x,t)\right]\psi_j$ with initial conditions $\psi_j(x,0)=\psi_j^{0}$, with $\psi_j^{0}$ being the eigenfunctions of the Schr\"odinger equation at initial time. This approach allows to describe by an exact solution the full dynamics after the quantum quench provided by the phase imprinting. In detail, we write the initial wavefunction as the groundstate of a ring potential in the presence of a barrier $\alpha\delta(x)$, constructed by the first $N$ single-particle orbitals  $\psi_j^{0}(x)$, which we then multiply by a phase profile that is induced by the phase imprinting, obtaining the initial wavefunction $\chi_j(x)=e^{2\pi i \ell x/L}\psi_j^{0}(x)$. The evolution is calculated by projecting such state with the eigenbasis of the unperturbed system $\psi_n(x,t)=\sum_{j}^{\infty}\langle\psi_{j}^{0}|\chi_n\rangle \psi_{j}^{0}(x)e^{-i\epsilon_j t/\hbar}$ and where $\epsilon_{n}$ is the $n$th single-particle eigenenergy
  \cite{Millard_1969,Wright_2002}.
The current within the TG regime, as being a local quantity, can be easily calculated using the occupations amplitudes of the single-particle eigenbasis, which in our case is found upon the projection over the initial state, as :
\begin{equation}
j(x,t)=\frac{\hbar}{m}\text{Im}\left[\sum_{n}^{\infty}f(\epsilon_n)\psi_n^{*}(x,t)\partial_x\psi_n(x,t)\right] \label{eq:current_temp_TG}
\end{equation}
with $f(\epsilon_n)$ being the Fermi-Dirac distribution. In this work we focus in particular on the average current flow along the ring, given by $J=\int dx j(x,t)/(LN)$.

\paragraph{Comparing different barriers}
We note that, in the thin-barrier limit, corresponding to $\sigma \ll n^{-1}$ and $\sigma\ll\xi$, where $\xi=\sqrt{\hbar^2/2mgn}$ is the healing length, the Gaussian and the delta barriers have a comparable effect and the strength of the delta barrier potential can be related to the parameters of Gaussian barrier through $V(x)=\alpha_{\rm eff}\delta(x)$ with $\alpha_{\rm eff}=\sqrt{2\pi}\sigma V_0$. This is useful to compare for example the dynamics obtained with the analytical two-mode model (with a delta barrier) to a full numerical simulation of the Gross-Pitaevskii equation (with a thin Gaussian barrier).

\section{\label{app:TMM}  Two-mode model for the Josephson oscillations of  the current}
In this section we obtain the equation of motion for the current as obtained from a  generalized version of the two-mode model widely used in Bose-Josephson junctions \cite{Smerzi1997}. This extension of the model also generalizes the two-mode approach presented in \cite{Ananikian_2006}, where all overlapping modes were real functions.

We start from an Ansatz for the condensate wavefunction:
\[
\Psi(x,t)=\phi_{1}(t)\varphi_{1}(x)+\phi_{2}(t)\varphi_{2}(x),
\]
where the mode orbitals $\varphi_{i}$, corresponding to  states carrying one positive/negative unit of angular momentum, fulfill the orthonormalization condition
\[
\int_{0}^{L}dx\,\varphi_{i}^{*}\varphi_{j}=\delta_{i,j},
\]
and where $\varphi_{1/2}$ are the $x$-coordinate representation of the wavefunctions built as a linear combinations of the first and second excited eigenfunctions $\psi_{i}$  of the non-interacting Hamiltonian (Eq.(1) of the main text with $g=0$, effects of interaction can also be included \cite{Nesterenko_2014}) as follows:
\[
\varphi_{1}(x)=\frac{(i\psi_1(x)+\psi_2(x))}{\sqrt{2}}\;, \quad  \varphi_{2}(x)=\frac{(i\psi_1(x)-\psi_2(x))}{\sqrt{2}}\;.
\]
The functions $\phi_{i}(t)$ are the time-dependent  amplitudes of populating the states $\varphi_{i}$. 


By introducing the previous Ansatz into the GPE equation, integrating over the spatial coordinate  and setting  {$\phi_{i}=\sqrt{J_i(t)}e^{i\theta_i(t)}$}
we obtain two coupled equations for  relative population $z=(J_1(t)-J_2(t))/J_T$ of the positive/negative angular momentum states and their relative phase
$\theta=\theta_2-\theta_1$:
\begin{eqnarray}
\dot{z} & = & -2\sqrt{\left(1-z^{2}\right)}\;\text{Im}\left[Ke^{i\theta}\right]+\left(1-z^{2}\right)|U_{2211}|\sin\left(2\theta-\arg\left(U_{2211}\right)\right)\nonumber\\
 & & +\sqrt{\left(1-z^{2}\right)}\left(1+z\right)|U_{2111}|\sin\left(\theta-\arg\left(U_{2111}\right)\right) +\sqrt{\left(1-z^{2}\right)}\left(1-z\right)|U_{2221}|\sin\left(\theta-\arg\left(U_{2221}\right)\right)
\end{eqnarray}
\begin{eqnarray}
\dot{\theta} & = & \left[E_{1}^{0}-E_{2}^{0}+\left(U_{1111}-U_{2222}\right)/2+z\left(U_{1111}+U_{2222}\right)/2-2z\Re\left[U_{1212}\right]\right] -z|U_{1122}|\cos\left(2\theta+\arg\left(U_{1122}\right)\right) \nonumber\\
 &  & +2K\cos\left(\theta\right)\frac{z}{\sqrt{\left(1-z^{2}\right)}}+\sqrt{\left(1-z^{2}\right)}\left[\left(|U_{1121}|\cos\left(\theta+\arg\left(U_{1121}\right)\right)-|U_{2122}|\cos\left(\theta+\arg\left(U_{2122}\right)\right)\right)\right] \nonumber\\
 &  & +\frac{z}{\sqrt{\left(1-z^{2}\right)}}\left[\left(z-1\right)|U_{1222}|\cos\left(\theta+\arg\left(U_{1222}\right)\right)-\left(z+1\right)|U_{2111}|\cos\left(\theta+\arg\left(U_{2111}\right)\right)\right].
\end{eqnarray}

The parameters entering the two-mode model are obtained from the mode orbitals according to:
\begin{gather*}
U_{i,j,k,l} = g_{1D}\int_{0}^{L}dx\,\varphi_{i}^*\varphi_{j}^*\varphi_{k}\varphi_{l},\\
K =-\int_{0}^{L}dx\,\left[\frac{\text{\ensuremath{\hbar}}^{2}}{2m}\varphi_{1}^{*}\partial_{x}^{2}\varphi_{2}+V(x)\varphi_{1}^{*}\varphi_{2}\right],
\\
E_{i}^{0} = \int_{0}^{L}dx\,\left[\frac{\text{\ensuremath{\hbar}}^{2}}{2m}\varphi_{i}^{*}\partial_{x}^{2}\varphi_{i}+V(x)|\varphi_{i}|^{2}\right].
\end{gather*}

\section{Multiple particle-hole excitations in the Tonks-Girardeau regime}

We detail here the calculation of the excitation frequencies and amplitudes  in the strongly interacting Tonks-Girardeau limit,  shown in Fig.~3(c) of the main text. We start rewriting Eq.~(2) for the current density following a phase imprinting according to 
\begin{align}
J=&\frac{\hbar}{Nm}\text{Im}\bigg[\sum_{k}^{\infty}\sum_{j}^{\infty} A_{j,k}  e^{-i(\epsilon_j - \epsilon_k)t/\hbar} \bigg],
 \label{eq:current_temp_TG_full}
\end{align}
where $\chi_n(x)=e^{-2\pi i l x/L}\psi_n(x)$ with $\psi_n$ being the eigenfunctions of the Schr\"odinger equation at initial time before the phase imprinting, and the  amplitude of the excitations is given by 
\begin{align}
A_{j,k} =& \frac{\hbar}{mL}\text{Im}\bigg[\sum_{n}^{\infty}f(\epsilon_n)\langle\chi_n|\psi_k\rangle \langle\psi_{j}|\chi_n\rangle \int_0^L dx\, \psi_{k}^{*}(x)\partial_x\psi_{j}(x) \bigg],
\label{eq:ampltude_excittaion}
\end{align}
with frequency of oscillation given by  $\omega_{j,k} = \frac{\epsilon_j - \epsilon_{k}}{\hbar}$. Figure 3(c) of the main text shows $\omega_{j,k}$ as a function of the barrier strength, with a greyscale color given by  $|A_{j,k}|$.

The  phase imprinting procedure  produces a highly excited state with respect to the system at rest, as all the particles start moving along the ring, going from a state with zero current to a circulating one. We  use the mapping of strongly interacting bosons onto non-interacting fermions to provide a microscopic insight in the dynamics of the current: Our analysis of the time dynamics of the mapped Fermi gas shows that   each initial non-moving single-particle state in the  Fermi sphere after the quench is projected onto  a superposition of doublet states with an energy splitting proportional to the gap opened by the barrier, leading to time oscillations of the current. Notice that since for weak barrier all the gaps are very close  in magnitude, the oscillations are almost synchronous for each particle, leading to coherent quantum phase slips, ie current oscillations with negligible damping. 

This picture shows that the quench is not a small perturbation on the system as the one giving rise eg to the dynamical structure factor; rather, multiple single-particle hole excitations occur and the whole Fermi sphere is involved in the dynamics. Figure~3(c) clearly shows that several excitations are involved.
One can also check what is the nature of each excitation:  as an example, the excitation with highest amplitude corresponds to the lowest energy single particle-hole excitation that can be created in our system, which in our particular case with an odd number of particles corresponds to a frequency of oscillation $\omega_{N+2,N+1}$. However, we observe that in fact, many excitations with lower associated energy are also excited. The existence of those excitations is possible because of the form of the  state just after the quench, which consists in a superposition of excited states of the unperturbed system, and not of a completely filled Fermi sphere.

The description of the dynamics is more complex for large barrier, since several excited states are also populated. Hence, even the dynamics of each  single particle state does not undergo to simple oscillations among two quantum states, and, furthermore, the oscillation frequency of the main mode differ from one particle to another one in the initial Fermi sphere. This provides a microscopic origin for the  dephasing of the current oscillations  observed in Fig.~3(a) of the main text. Furthermore, at finite temperature the coherent processes of each individual particle, labeled by $n$ in \eref{eq:ampltude_excittaion}, are mixed by the Fermi distribution $f(\epsilon_n)$, giving rise to, as is also known in previous works, incoherent phase slips and effective damping in Fig.~3(b).

\bibliography{biblio.bib}

\clearpage